\documentclass[final,1p]{elsarticle}

\makeatletter
\def\ps@pprintTitle{%
 \let\@oddhead\@empty
 \let\@evenhead\@empty
 \def\@oddfoot{\centerline{\thepage}}%
 \let\@evenfoot\@oddfoot}
\makeatother

\usepackage{graphicx}
\usepackage{amssymb}
\usepackage{amsthm}
\usepackage{bm,amsmath,amssymb,amsfonts}

\usepackage{lineno}
\usepackage{siunitx}
\biboptions{sort&compress,square}

\usepackage{tabularx}
\usepackage{booktabs}

\usepackage[hyphens]{url}

\newcommand{\eqn}[1]{Eq.~(\ref{#1})}
\newcommand{\eqns}[2]{Eqs.~(\ref{#1}) and (\ref{#2})}

\newcommand{\fig}[1]{Fig.~\ref{#1}}

\newcommand{\sect}[1]{Sec.~\ref{#1}}

\newcommand{\tableref}[1]{Table~\ref{#1}}

\renewcommand{\b}[1]{\boldsymbol{#1}} 
\renewcommand{\o}[1]{\overline{#1}}

\newcommand{\Grad}{\ensuremath{\mbox{Grad}}}
\newcommand{\Div}{\mbox{Div}}

\renewcommand{\d}{\mathsf{d}}

\newcommand{\dfracp}[2]{\dfrac{\partial #1}{\partial#2}}

\usepackage{color}
\newif\ifproofread

\newcommand{\changemarker}[1]{%
\ifproofread
#1%
\else
\textcolor{blue}{#1}%
\fi
}

\journal{}

\begin{document}

\proofreadtrue

\begin{frontmatter}


\title{Microstructurally-based constitutive modelling of the skin -- Linking intrinsic ageing to microstructural parameters}

\author[label1]{D. Pond}
\author[label1,label2,cor1]{A.T. McBride}
  \ead{andrew.mcbride@glasgow.ac.uk}
\author[label3]{L.M. Davids}
\author[label1]{B.D. Reddy}
\author[label4,label5]{G. Limbert}


 \address[label1]{Centre for Research in Computational and Applied Mechanics, University of Cape Town, South Africa}
 \address[label2]{School of Engineering, University of Glasgow, G12 8QQ, United Kingdom}
 \address[label3]{Department of Medical Biosciences, University of the Western Cape, South Africa}
 \address[label4]{national Centre for Advanced Tribology at Southampton (nCATS) / Bioengineering Science Research Group, Faculty of Engineering and the Environment, University of Southampton, United Kingdom}
 \address[label5]{Laboratory of Biomechanics and Mechanobiology, Department of Human Biology, Faculty of Health Sciences, University of Cape Town, South Africa}
 \cortext[cor1]{Corresponding author.}



\begin{abstract}
A multiphasic constitutive model of the skin \changemarker{that implicitly accounts for the process of intrinsic (i.e.\ chronological) ageing via variation of the constitutive parameters} is proposed.
The structurally-motivated constitutive formulation features distinct mechanical contributions from collagen and elastin fibres.
The central hypothesis underpinning this study is that the effects of ageing on the mechanical properties of the tissue are directly linked to alterations in the microstructural characteristics of the collagen and elastin networks.
Constitutive parameters in the model, corresponding to different ages, are identified from published experimental data on bulge tests of human skin.
The identification procedure is based on an inverse finite element method.
The numerical results demonstrate that degradation of the elastin meshwork and variations in anisotropy of the collagen network are plausible mechanisms to explain ageing in terms of macroscopic tissue stiffening.
Whereas alterations in elastin affect the low-modulus region of the skin stress-strain curve, those related to collagen have an impact on the linear region.
\end{abstract}


\end{frontmatter}



%

\section{Introduction} \label{sec_introduction}

The skin acts as a  biophysical interface between the human body's internal and external environments.
Due to this critical role, significant research has been undertaken to elucidate the complex processes underlying the skin's physiology and biophysics \citep{Jor2013, Li2015, Limbert2017, Silver2003}.
Of particular relevance to the present study, is the search for plausible microstructural mechanobiological mechanisms responsible for alterations of the skin’s macroscopic mechanical properties with age.

Between 2001 and 2011, the UK population aged 65 and over increased by 0.92 million \citep{web2010}.
The UK population aged 85 and over increased by almost 25\% (from 1.01 million to 1.25 million).
By 2039, almost a quarter of the UK population will be 65 and over \citep{web2016}.
As the aged population continues to increase, two main issues emerge.
First, the need for medical treatment tends to increase with longevity.
Second, the ageing process itself results in significant degradation of the biophysical properties of skin.
This leads to further, and potentially life-threatening, complications such as skin tears and associated chronic wounds \citep{Morey2007}.

The skin accounts for up to 16\% of an adult’s total body weight and covers an average surface area of about \SI{1.6}{\m^2} \citep{Shimizu2007}.
Given this, and the role of the skin as the prime line of defence against the external environment, it is clear that any significant alteration of mechanical properties that could compromise the structural integrity and barrier function, has the potential to cause serious detrimental health issues.
Moreover, both in the UK and globally, the ageing population is rapidly becoming a significant market segment across many industrial sectors, including medical devices, consumer goods, personal care products, sports equipment and consumer electronics.
It is therefore important to understand how age-related alterations of skin biophysics can be accounted for in the development of new or improved products.

\begin{figure}[htb!]
    \centering
    \includegraphics[width=\textwidth]{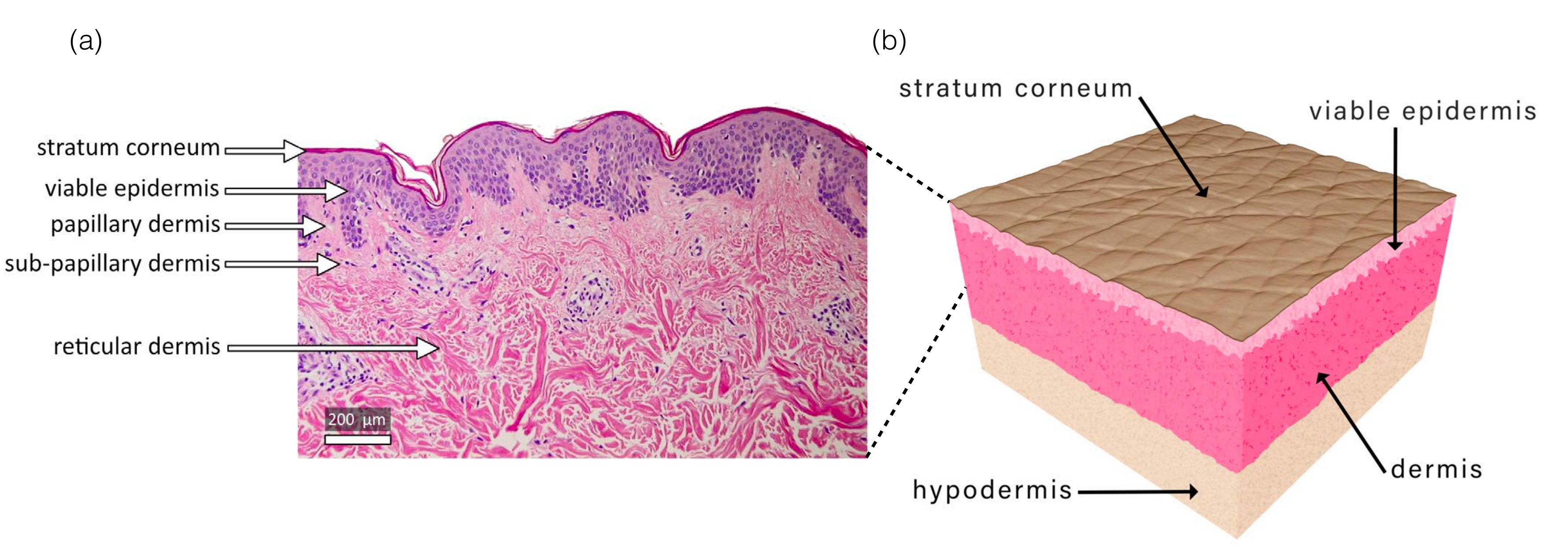}
    \caption{(a) Histological section of haematoxylin and eosin stained back human skin sample obtained from a 30 years-old healthy white caucasian female volunteer following biopsy (10 x magnification, image resolution: 1600 x 1200 pixels, imaged using a modified Nikon E950 camera).
    Image courtesy of Dr.\ Maria-Fabiola Leyva-Mendivil (University of Southampton, UK); (b) Schematic representation of the mesoscopic multi-layer structure of the skin.}
    \label{fig_skin_structure}
\end{figure}

In accordance with the universal principle of close integration of structure and function in biological tissues \citep{Fung1981}, the human skin features a complex multiscale structural architecture.
For simplicity's sake, the skin is often mesoscopically described as a three-layer structure made of an epidermis, dermis and hypodermis (see \fig{fig_skin_structure}(b)) \citep{Shimizu2007, Limbert2014, Lanir1987}.

The dermis can be decomposed into three main layers: the papillary layer juxtaposed to the epidermis, the sub-papillary layer underneath, and the reticular layer which is connected to the underlying subcutaneous tissue.
The subcutaneous tissue is the layer between the dermis and the fascia which is a band of connective tissue, primarily collagen, that attaches, stabilises, encloses, and separates muscles and other internal organs.
The thickness of the subcutaneous tissue is highly variable, both intra- and inter-individually.
This layer is mainly composed of adipocytes.
Its role is to provide mechanical cushioning, heat generation, insulation, and as a reserve of nutrients.

Like any other organ of the human body, the skin inexorably undergoes what is termed chronological ageing, or more correctly, intrinsic ageing: a series of biochemical molecular degenerative changes occurring as the indirect result of the passage of time and progression into older age.
These alterations involve decreased proliferative capacity which leads to cellular senescence and altered biosynthetic activity of skin-derived cells.
Deterioration of the skin barrier function, a reduced rate of turnover of epidermal cells (e.g.\ keratinocytes), alteration of the biochemical and mechanical properties of the collagen network in the dermis, macroscopic stiffening, loss of elasticity and a poor vascular network around hair bulbs and glands are manifestation of these intrinsic ageing-induced physiological changes \citep{Ramos2007}.
The general consensus is that intrinsic ageing is triggered by two main mechanisms which can operate in concert: DNA damage and chromosomes’ telomere shortening \citep{Assaf2014, Naylor2011, Gilchrest1992, Goukassian2000, Giacomoni2004}.
An important aspect to consider is that intrinsic ageing typically occurs in combination with extrinsic ageing which is the result of external environmental and lifestyle factors, particularly exposure to ultraviolet (UV) radiations from sun light \citep{Fisher1997, Berneburg2000, Kligman1969}, and increasingly, sunbeds \citep{Fisher2002, Diffey2003}, as well as smoking \citep{SandbyMoller2003, Vierkotter2012} and air pollution \citep{Vierkotter2012, Benedetto1998}.
Extrinsic ageing due to UV radiation exposure is called photoageing \citep{Berneburg2000, Kligman1969}.

A popular view on skin ageing is embodied by the microinflammatory model advocated by \citet{Giacomoni2004}.
These authors proposed a mechanistic biochemical model of ageing which is not rooted in any mathematical formulation but is rather a descriptive qualitative model integrating information about the ageing of the extracellular matrix and that of the dermis.
The model accounts for both intrinsic and extrinsic ageing effects through loss of elasticity, resilience and flexibility of the dermis, and also integrates the formation of wrinkles and the thinning of the epidermis.
Ageing affects both structure and function of the skin and these alterations have important consequences for its physiology \citep{Assaf2014, Naylor2011, Pawlaczyk2013, Sherratt2013}, rheology \citep{Silver2002, Ruvolo2007}, tribology \citep{LeyvaMendivil2017, LeyvaMendivil2017a, LeyvaMendivil2015} and the likelihood of developing particular age-related skin disorders such as xerosis and pruritus \citep{Hahnel2017} or skin tears \citep{Morey2007}.

As a result of biochemical changes associated with ageing, the mechanical behaviour of the skin is significantly altered through time-dependent variations in the structural and mechanical properties of its elemental constituents (e.g.\ proteoglycans, collagen, elastin and keratin).
The effects of skin ageing become apparent in the form of skin wrinkles - particularly on the face and hands - which are often unconsciously or consciously used as criteria to evaluate age.
These morphological changes of the skin surface are manifestations of complex coupled biophysical phenomena where mechanics is believed to play a critical role as in many other remodelling and morphomechanical processes \citep{Goriely2007, Ambrosi2011}.
Unravelling the inherent complexity of the skin ageing process, firstly by identifying its biophysical drivers, underlying factors and effects, and secondly, by gaining a mechanistic insight into their interplay, is a formidable challenge at both experimental and modelling levels.
However, building upon the extensive body of work on the constitutive modelling of the skin \citep{Jor2013, Li2015, Limbert2017}, there are opportunities to extend existing models and develop mechanistic constitutive formulations that could first describe, and ultimately, predict ageing effects.
These hypothesis-driven research tools have the potential  to unveil the biophysical complexity of skin physiology as well as mechanobiological aspects associated with diseases and the ageing process.

Various phenomenological and structurally-based mechanical models have been proposed and adopted to model the skin at finite deformations.
For a good coverage of the relevant literature see the monograph by \citet{Xu2011}, review papers by \citet{Jor2013, Li2015, Limbert2017} or book chapters by \citet{Flynn2014, Limbert2014}.
Structurally-based models attempt to reflect the contributions and mutual interactions from the primary constituents of the dermal layer, i.e.\ collagen, elastin and ground substance.
Developing these type of models is desirable as they offer the ability to link the microscopic constituent characteristics (i.e.\ materials and structures) to the macroscopic response of the tissue \citep[see][]{Kuhl2005, Kuhl2007, Garikipati2004}.
\citet{Mazza2005, Mazza2007} developed a non-linear constitutive model to simulate ageing of the human face.
The elasto-visco-plastic constitutive model is based on the formulation by \citet{Rubin2002}.
\citet{Mazza2005, Mazza2007} extended the model of \citet{Rubin2002} by including an ageing parameter equipped with its own time evolution equation.
 This ageing-driven parameter was a modulator of tissue stiffness.
 A four-layer model of facial skin combined with a face-like geometrical base was developed.
The work highlighted the utility of the model to study the effects of skin ageing on facial appearance.
\citet{Maceri2013} proposed an age-dependent multiscale mechanical model for arterial walls that effectively coupled elastic nanoscale mechanisms, linked to molecular and cross-link stretching, to micro- and macroscopic structural effects.
The model successfully captured the age-dependent evolution of arterial wall mechanics through alterations of the constitutive parameters including geometric characteristics of collagen fibres, cross-link stiffness of collagen fibrils and volume fraction of constituents.

To date, however, no mechanistic constitutive model for the skin has been developed that is capable of simultaneously capturing intrinsic ageing through evolution of  material and structural constitutive parameters whilst being embedded in the rigorous framework of non-linear continuum mechanics.
The goal of the research presented here is to develop an experimentally-based mathematical and computational model of the skin to study the interplay of its material and structural properties and their evolution as a result of the intrinsic ageing process.
\changemarker{Here, rather than proposing evolution laws governing the alteration of constitutive parameters as a function of ageing, we conduct an experimentally-based parametric study of skin's mechanics to investigate the correlation between the process of ageing and the constitutive model's parameters.}
Plausible mechanisms associated with ageing-induced material and microstructural evolution are explored in an attempt to explain observed effects associated with ageing (e.g.\ macroscopic stiffening of the tissue).

The paper is organised as follows.
The general structural and material properties of the skin are discussed in \sect{sec_struct_mat_props}.
Special focus is placed on the evolution of these characteristics with intrinsic ageing.
A continuum-mechanics description of the skin is presented in \sect{sec_continuum_model_skin}.
\sect{sec_micro_modelling_age} is concerned with the formulation of hypotheses on the link between ageing and the evolution of material and structural properties and their effects on the observed macroscopic mechanical response of the skin.
Experimental bulge test data \citep{Tonge2013, Tonge2013b} are used in \sect{sec_FE_bulge_test} to calibrate the finite element approximation of the continuum model.
The proposed ageing model is then used to predict the influence of ageing on the mechanical response of the skin in \sect{sec_validation_age_model}.
\sect{sec_discussion} includes a discussion of the results and conclusions.

\section{Structural and material properties of the ageing skin} \label{sec_struct_mat_props}

\subsection{Structural aspects}

It is generally accepted that the bulk of the skin’s response to loading is due to the dermal layer \citep{Silver2001, Silver2002}.
The dermis, is composed of papillary, sub-papillary and reticular layers, and is comprised of a matrix of collagen, elastin, other elastic fibres and ground substance.
The expression of the macroscopic mechanical properties of the skin is due to these components, their structural organisation and their mutual interactions.
The ground substance is a gel-like amorphous phase mainly constituted of proteoglycans and glycoproteins (e.g.\ fibronectin) as well as blood and lymph-derived fluids which are involved in the transport of substances crucial to cellular and metabolic activities.
Proteoglycans are composed of multiple glycosaminoglycans (i.e.\ mucopolysaccharides) interlaced with back bone proteins.
Dermal fibroblasts produced glycosamine which is rich in hyaluronic acid and therefore plays an essential role in moisture retention.

Collagen has been found to make up approximately 66-69\% of the fractional volume of the dermis \citep{Silver2002}, and approximately 34\% \citep{Leveque1980} and 70-80\% \citep{Reihsner1995} of the wet weight and dry weight of the skin, respectively.
Experiments where collagen was isolated through enzymatic treatment \citep{Oxlund1980, Oxlund1988, PaillerMattei2008}, conclude that collagen is responsible for the tensile strength of the skin.
The papillary layer (\fig{fig_skin_structure}(a)) is defined by the rete ridges protruding into the epidermis and contains thin collagen fibres, sensory nerve endings, cytoplasms and a rich network of blood capillaries.
The sub-papillary layer - the zone below the epidermis and papillary layer - features similar structural and biological components to those of the papillary layer.

Besides the dominant content of types I and III collagen (respectively 80\% and 15\% of total collagen content), the reticular layer is innervated and vascularised, contains elastic fibres (e.g.\ elastin) and the dermal matrix made of cells in the interstitial space.
Cells present in the reticular dermis include fibroblasts, plasma cells, macrophages and mast cells. Collagen fibres in the papillary and sub-papillary dermis are thin (because of their low aggregate content of fibrils) and sparsely distributed while reticular fibres are thick, organised in bundles and densely distributed.
Fibrils are typically very long, 100 to 500 \si{\nano\meter} in diameter featuring a cross striation pattern with a 60 to 70 \si{\nano\meter} spatial periodicity.
 The diameter of thick collagen bundles can span 2 to 15 \si{\micro\meter}.
 Birefringence techniques have been used to characterise the orientation of a supramolecular organisation of collagen bundles in skin \citep{Ribeiro2013}.

Contributing approximately 2-4\% of the dry weight of skin \citep{Tonge2013}, elastin fibres are highly compliant with the ability to stretch elastically to twice their original length \citep{Gosline2002}.
Their diameter ranges between 1 and 3 \si{\micro\meter}.
Their mechanical intertwining with the collagen network of the dermis is what gives the skin its resilience and recoil ability.
This is evidenced by the correlation between degradation of elastin/abnormal collagen synthesis associated with ageing and the apparent stiffening of the dermis \citep{Sherratt2013}.
\changemarker{This aspect will be shown to be particularly important for the mechanistic model linking ageing to the alteration of the skin's constitutive model parameters presented in \sect{sec_micro_modelling_age}.}
The diameter of elastic fibres in the dermis is inversely proportional to their proximity to the papillary layer where they tend to align perpendicular to the dermal-epidermal junction surface.

\subsection{Material aspects}

At a macroscopic level, the mechanical properties of the skin are anisotropic and inhomogeneous.
This is due to the complex hierarchical structure and materially non-linear constituents and residual tension lines in the skin (i.e.\ the so-called ``Langer lines'', first recognised by the Austrian anatomist Karl Langer in his seminal study \citep{Langer1978}).

The angular distribution of collagen fibres and the non-uniform fibre geometry means that under stretch not all fibres straighten and stretch equally.
This accounts for the anisotropic stiffness response when load is applied either along or across the preferential fibre direction.
The magnitude of these directional effects has been the subject of several modern studies.
The Young’s modulus parallel to the Langer lines was found to be greater than that perpendicular by a ratio of approximately 2.21:1 \citep{Lapeer2010}.
Similarly, \citet{Reihsner1995} found that the degree of anisotropy differs across anatomical site, an observation also made by Langer in his original study.
In addition, they found that in situ stress ranges from 0.2-1.6 \si{\N\per\mg} along the Langer lines and 0.1-1.3 \si{\N\per\mg} perpendicularly to them, with the degree of anisotropy differing between principle stress components by 0.1-0.3 \si{\N\per\mg}.

When the skin is stretched, the elastin fibres are the first to take on stretch \citep{Silver1992}, indicating that this contribution is important at low strain levels.
Stress-strain curves of elastin-free skin \citep{Oxlund1988} show that elastin supports the entire load up to 50\% strain after which the strength rapidly increases due to the collagen.
The elastic modulus of elastin has been found to be around \SI{1}{\MPa}.
This agrees with the Young’s modulus of skin at low strain.
In addition, elastin is not strong enough to provide much tensile strength at higher strains \citep{Tregear1969}.
\citet{Reihsner1995} state that elastin is responsible for the recoiling  of the skin and collagen after stress is applied.
Following degradation of the elastin through the use of elastase, \citet{Oxlund1988} found that the large strain response occurs sooner for a given tensile load.
This seems to suggest that, in the absence of elastin, collagen fibres take on load at lower strain levels than when elastin is present.
The ground substance has been shown to influence mainly  the viscoelastic properties of the skin because of high-water content and complex time-dependent interstitial fluid motion.
Upon removal of various macromolecules within the ground substance, \citet{Oxlund1980} showed that there was no effect on the mechanical response of rat skin, while \citet{Oomens1987} suggests that ground substance probably only plays a \changemarker{major role} when soft tissue is subject to compression.

\begin{figure}[htb!]
    \centering
    \includegraphics[width=0.8\textwidth]{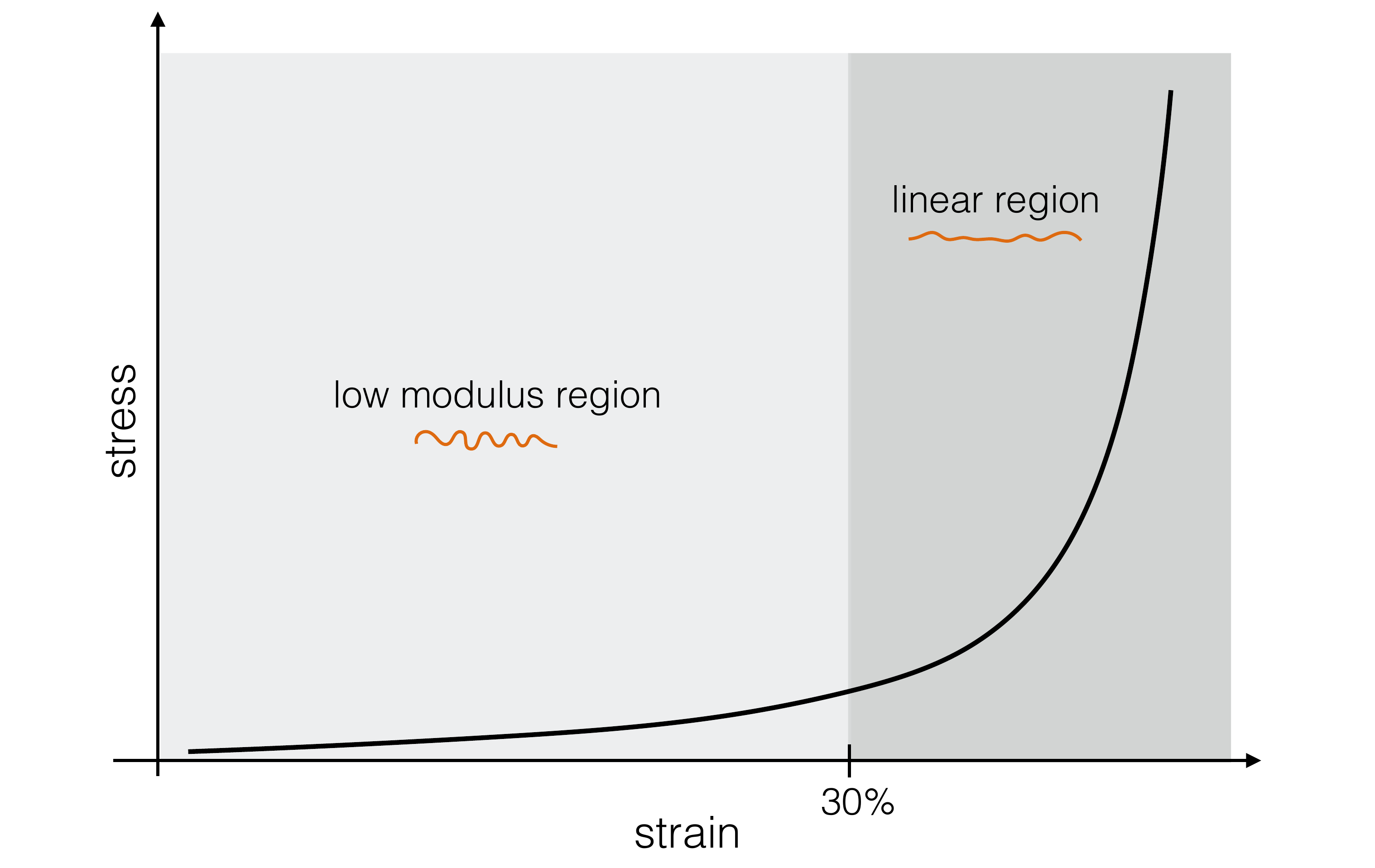}
    \caption{Typical stress-strain curve for skin subjected to uniaxial extension.}
    \label{stress_strain_skin}
\end{figure}

\fig{stress_strain_skin} highlights the typical strain hardening characteristics of skin subjected to uniaxial extension where each portion of the stress-strain curve can be related to the dermal constituents as described in the next paragraph.

\emph{Low modulus portion of the stress-strain curve:}
 this is associated with the gradual straightening of crimped collagen fibres. During this stage, the greatest resistance to loading is generated by the elastin and ground substance, with collagen fibres offering very little resistance. The low modulus portion can be further divided into two phases:
\begin{enumerate}[leftmargin=12pt]
  \item[Phase 1:] wavy collagen fibres are still relaxed and elastin fibres take on the majority of the load.
  \item[Phase 2:] collagen fibres start to uncrimp, then elongate and eventually start to bear load.
\end{enumerate}

\emph{Linear region of the stress-strain curve:}  collagen fibres straighten and align with the load direction. Straightened collagen fibres strongly resist loading. This results in the rapid stiffening of the skin. The steep linear stress-strain relation is due to stretching and slippage between fibrils and molecules.

\emph{Final yield region of the stress-strain curve:}
tensile strength of collagen fibres is reached and fibres begin to sequentially break (not shown in \fig{stress_strain_skin}).

\subsection{Variations of skin structural and material properties with age}

It is generally accepted that skin thickness decreases with age \citep{PaillerMattei2008, Escoffier1989}.
\citet{Pawlaczyk2013} found that there is an overall loss of 0.7-0.8 \si{\mm} of thickness in older skin.
It has been found that skin thickness reaches a maximum around the fourth decade for men and third decade for women after which there is a gradual decrease \citep{Leveque1980}, a result similar to that established by \citet{Diridollou2001} who found that after an initial increase during maturation (0-20 years), thickness remains constant to about the age of 60 followed by a decrease as follows
\begin{equation}\label{t_with_time}
t = - 6 \times 10^{-3}~[\text{mm/year}] \times \text{age}~[\text{years}] + 1.3~[\si{\mm}] \, .
\end{equation}

Moreover, the rate of decrease is more significant in female subjects.
It is important to highlight that these observations on skin thickness apply to the skin as a whole composite structure as individual skin layers may follow different trends depending on whether intrinsic and extrinsic ageing effects are considered separately or assumed to be combined.
After 20 years of age, across all layers, the skin thickness starts to diminish at a rate that increases with age \citep{Oriba1996}.
Between 30 and 80, the unexposed skin can lose up to 50\% of its thickness and this effect is accentuated in zones exposed to sunlight such as the face or neck. Overall, epidermal thickness drops by about 6.4\% per decade faster in women than men.
 It is generally believed that the reported reduction in dermal thickness is mainly caused by the loss of dermal collagen and elastin in elderly adults \citep{Duncan1997}.
 It was shown that in post-menopausal women a 1.13\% per year skin thickness reduction is correlated with a 2\% decrease per year in collagen content \citep{Brincat1987}.
 It is also accepted that skin ageing is characterised by an increase in macroscopic or apparent stiffness \citep{Pawlaczyk2013, Xu2011, Alexander2006}, although there is little agreement on the magnitude of the ground state's Young’s modulus, or the age of onset of stiffening.

As mentioned by \citet{Xu2011}, there is a sudden increase in the Young’s modulus of the skin at age 30 of around 50\%, whereas others quote an increase from the age of 45.
It was observed by \citet{Escoffier1989} that there is an increase of around 20\% after the age of 70, which is backed by the findings of \citet{Leveque1980}.
Furthermore, \citet{Alexander2006} found that the stiffness of skin starts increasing from the  age of 25 but noted  that the variation in results increases with age.
This suggests that the process of skin ageing is a highly patient specific and may explain the large variation in results reported in the literature.
It was found by \citet{Escoffier1989} and \citet{Leveque1980} that intrinsic skin extensibility (i.e.\ a standardised mean extensibility to account for varying skin thickness) decreases with age, while \citet{Alexander2006} found that intrinsic extensibility decreases by around 35\% after the age of 65 which agrees with the findings of \citet{Xu2011} in that maximum skin elongation is found between ages 35 and 55. Similarly, skin elasticity (i.e.\ recoil ability) decreases with age \citep{Xu2011, Escoffier1989}, which was similarly found by \citet{Henry1997}, but this may include the effects of UV radiation.

It has been widely observed that, with age, there is a decrease in the initial portion of the elongation-stress curve \citep{Reihsner1995, Alexander2006, Daly1979} which means that the onset of stiffening associated with recruitment of collagen fibres occurs at lower stretch.
As elastin is primarily responsible for the low stretch response, it is clear that there must be some modification to the elastin network with age.
Elastin is a highly stable protein, the biosynthesis of which remains steady for the first 40 to 50 years, after which there is a decline.
Various authors \citep{Silver2002, Reihsner1995, Alexander2006, Daly1979} attribute the decrease in the low-modulus portion to the gradual degradation of the elastin network, as well as the deposition of amorphous elastin.
This statement is backed by the observation that, upon enzymatic removal of elastin from skin samples, similar effects were obtained to those of ageing.

In the linear region of the stress-strain curve (see \fig{stress_strain_skin}), collagen fibres take on tension which results in the typical nonlinear response.
It has been reported that the slope of the linear portion of the elongation-stress curve tends to increase with age \citep{Alexander2006}.
This suggests an apparent stiffening of the collagen fibre network with age but not necessarily a stiffening of the collagen fibres themselves.
This aspect is supported by the experimental evidence obtained by \citet{Daly1979} and \citet{Reihsner1995} who note that the final slope of the strain-stress curve of skin in tension remains constant with age, and that the stiffness of the collagen remains constant, possibly due to the reduction in the collagen content in the dermal layer \citep{Chung2000, Jenkins2002}.
There is therefore suggestion that the alteration of the skin apparent stiffness is not due to a stiffening of the collagen fibres, but rather to an alteration in the structure of the collagen network.
It was found that alterations in total skin thickness were proportional  to the alterations in dermal thickness and that, due to the compacting of the dermis with age, collagen bundles tend to flatten and unravel \citep{Leveque1980}.
Coupled with the loss of elastic integrity, \citet{Diridollou2001} postulate that, due to the dermal thinning, collagen fibres may tend to unfold to some extent, which could further explain the reduction in skin extensibility and the smaller low-modulus portion, as collagen is activated sooner.

\citet{Agache1980, Escoffier1989, Reihsner1995} mention that there is increased crosslinking with age, which would reduce any slippage between neighbouring fibres and stiffen up the collagen network. Anisotropy tends to increase with age suggesting that there is an increase in alignment along Langer lines with age \citep{Ruvolo2007, Vexler1999}, or at least, a correlation between age, microstructurally-induced anisotropy and Langer lines, although \citet{Tonge2013} found a decrease in overall anisotropy with age.

In summary, three factors that certainly influence biomechanical behaviour of skin with age are:
\begin{enumerate}
  \item Increased cross-linking between collagen fibres - collagen network density increases with age as well as a decrease in fibre free length. Collagen fibres are thus more compact and appear to unravel.
  \item The loss of elasticity at low strain is attributed to the destruction of the elastin network. Elastic fibres become thinner and fractionated with age. The decreased extensibility properties of skin with age are generally attributed to the loss of elastin in the upper dermis with age.
  \item Water content tends to decrease with age, which alters the overall viscoelastic properties.
\end{enumerate}

\section{Continuum model of skin} \label{sec_continuum_model_skin}

\subsection{Notation}
Direct notation is adopted throughout.
The scalar product of two vectors $\b{a}$ and $\b{b}$ is denoted by $\b{a} \cdot \b{b}$.
The scalar product of two second-order tensors $\b{A}$ and $\b{B}$ is denoted by $\b{A} : \b{B}$.
The composition of two second-order tensors $\b{A}$ and $\b{B}$ is denoted by $\b{A}\b{B}$.
The action of a second-order tensor $\b{A}$ on a vector $\b{b}$ is a vector denoted by $\b{A}\b{b}$.
The unit basis vectors in the Cartesian (standard-orthonormal) basis are \{$\b{e}_1, \b{e}_2, \b{e}_3$\}.

\subsection{Finite strain kinematics}

Consider a continuum body, here representing skin, defined in relation to the Cartesian reference frame.
The reference configuration is defined as the placement of this body at time $t=0$, with that region denoted by $\Omega_0$ and boundary $\partial\Omega_0$ with outward unit normal $\b{N}$.
 As the body deforms, this region takes on subsequent configurations.
 At a current time $t$, the body occupies \changemarker{the region} $\Omega$ referred to as the current configuration and current boundary surface $\partial\Omega$ with outward unit normal $\b{n}$.
 For an extensive overview of continuum mechanics the reader is referred to \citep{Holzapfel2000Book, Marsden1984}, among others.

 It is assumed that there exists a mapping $\b{\chi}$ such that each material point with position $\b{X} \in \Omega_0$ uniquely maps to a spatial point with position $ \b{x} \in \Omega$ at time $t$, i.e.\
 \begin{equation*}
   \b{x} = \b{\chi}(\b{X},t) \qquad \forall \b{X} \in \Omega_0 \, .
 \end{equation*}

 The motion $\b{\chi}$ is assumed to possess continuous derivatives both in space and time.
 \changemarker{The deformation of the body can be characterized} by the deformation gradient $\b{F}$, i.e.\ the material gradient of the motion, defined by
\begin{equation*}
  \b{F}(\b{X},t) := \dfrac{\partial\b{\chi}(\b{X},t)}{\partial \b{X}} =\Grad \b{\chi}(\b{X},t) \, .
\end{equation*}
The determinant of $\b{F}$ is defined by $J := \det \b{F} > 0$.
The right Cauchy--Green tensor $\b{C}$, defined by
\begin{equation*}
  \b{C} := \b{F}^T\b{F} \, ,
\end{equation*}
provides a stretch measure in the reference configuration.
Additionally, the \changemarker{principal} scalar invariants of $\b{C}$ are defined by
\begin{align}\label{I1_to_I3}
I_1(\b{C}):=\text{tr}(\b{C}) = \b{I}:\b{C} \, , &&
I_2(\b{C}):=\dfrac{1}{2} \left(\text{tr}(\b{C})^2-\text{tr}(\b{C}^2)\right) \, , &&
I_3(\b{C}):=\text{det}(\b{C}) \, .
\end{align}

Consider now the case of transverse isotropy where the material properties depend on a single given direction  (which is a defining feature of the present constitutive formulation for skin).
The preferred material direction is given by $\b{v}_0$.
The fabric tensor is defined by
\begin{align*}
  \b{A}_0 := \b{v}_0 \otimes \b{v}_0 \, .
\end{align*}
An additional invariant characteristic of transverse isotropic symmetry is defined by
\begin{align}\label{I4}
  I_4(\b{C}, \b{v}_0) = \b{v}_0 \cdot \b{C}{v}_0 = \lambda^2,
\end{align}
where $\lambda$ is the principal stretch along vector $\b{v}_0$ defined in the reference configuration.

 \changemarker{Note, five scalar-valued tensor invariants are necessary to form the irreducible integrity bases of the tensors $\b{C}$ and $\b{v}_0 \otimes \b{v}_0$ \citep{Boehler1978, Spencer1992}.
 Thus, to fully describe transverse isotropic symmetry, an additional invariant $I_5$ must be introduced \citep{Boehler1978, Holzapfel2000Book, Limbert2002, Spencer1992}, where
 \begin{align*}
   I_5 := \b{v}_0 \cdot \b{C}^2{v}_0 \, .
 \end{align*}
The fifth invariant is often discarded for mathematical convenience and because of the intrinsic difficulty of providing it a direct physical interpretation that can be linked to  measurement \citep{Schroder2003}.
\citet{Destrade2013} demonstrated that excluding $I_5$ as an argument of the strain energy function of a transversely isotropic hyperelastic material could result in non-satisfaction of kinematic
constraints present in physical tests.
As is often the case in constitutive models of biological soft tissues, and for simplicity's sake with regards to the interpretation of the results of our study, the fifth invariant $I_5$ is not considered here.
For a general discussion on the inclusion of additional invariants to capture the mechanical response of transversely isotropic materials, see \citep{Destrade2013}.}

\subsection{Constitutive relations for invariant-based transversely isotropic hyperelasticity}

A hyperelastic material is one for which a free energy $\psi$, defined per unit reference volume, acts as a potential for the stress.
For homogeneous materials, the free energy is purely a function of the deformation gradient, i.e.\ $\psi = \psi(\b{F})$,  \changemarker{and of any additional tensor agency (e.g.\ structure tensors).}
\changemarker{Note that, from standard objectivity arguments arguments, $\psi$ depends on $\b{F}$ through $\b{C}$ \citep[see e.g.][]{Holzapfel2000Book, Marsden1984}.}
\changemarker{As a general procedure to formulate constitutive equations for hyperelastic materials, one can postulate the existence of a strain energy density $\psi$, that is an isotropic function of its deformation or strain-invariant arguments.}
The first and second Piola--Kirchhoff stress measures, $\b{P}$ and $\b{S}$, are defined by
\begin{align*}
  \b{P} = \dfracp{\psi(\b{F})}{\b{F}}
  && \text{and} &&
  \b{S} = 2\dfracp{\psi(\b{C})}{\b{C}} \, ,
\end{align*}
where $\b{P} = \b{F} \b{S}$.
\changemarker{Under the assumption that the material modelled exhibits some form of material symmetry, the dependence of the free energy on $\b{C}$ can be be expressed in terms of the $n_\text{inv}$ invariants of $\b{C}$.}
The second Piola--Kirchhoff stress is thus given by
\begin{align}
  \b{S} = 2 \sum_{i=1}^{n_\text{inv}} \dfracp{\psi(\b{C})}{I_i} \dfracp{I_i}{\b{C}} \, . \label{PK2}
\end{align}

\subsection{Balance relations}

\changemarker{The balance of linear momentum, in the absence of inertial or body forces, and the natural boundary condition are given by}
\begin{align}
  \Div \b{P} = \b{0} && \text{in } \Omega \, , \label{balance_momentum} \\
  \b{T} = \b{P} \b{N} =  \o{\b{T}} && \text{on } \partial \Omega_{0,\text{N}} \, , \label{neumann_bc}
\end{align}
\changemarker{where $\Div$ is the material divergence operator.}
The Piola traction $\b{T}$ is prescribed on the Neumann part of the boundary $\partial \Omega_{0,\text{N}} \subset \partial \Omega_0$.
Dirichlet boundary conditions on the motion $\b{\chi} = \o{\b{\chi}}$ are prescribed on $\partial\Omega_{0,\text{D}}$ where $\partial \Omega_0 = \partial\Omega_{0,\text{D}} \cup \partial\Omega_{0,\text{N}}$ and $\partial\Omega_{0,\text{D}} \cap \partial\Omega_{0,\text{N}} = \emptyset$.

\subsection{Constitutive assumptions for the dermis} \label{sec_constitutive_dermins}

Based on the experimental evidence reported in \sect{sec_struct_mat_props}, it is assumed that elastin and ground substance are the main contributors to the low modulus portion of the stress-strain curve (see \fig{stress_strain_skin}) which is largely linear and isotropic.
The second region of the loading curve is dominated by the mechanical response of collagen fibres.
As the collagen fibres straighten and take on load, they exponentially resist further stretch which results in a rapid nonlinear strain stiffening or locking behaviour. This behaviour is also strongly anisotropic due the inherent preferred alignment of collagen fibres along the direction of extension.
As the stiffening response of the skin is dominated by the response of the collagen network, the formulation of an effective microstructurally-motivated model is essential.
Hence, it is postulated that the free energy describing the overall mechanical behaviour of the dermis, assumed to be that of the skin because of the negligible contributions of the epidermis in tension, is given by
\begin{align} \label{psi}
  \Psi = \Psi_\text{gs} + \Psi_\text{elastin} + \Psi_\text{collagen},
\end{align}
where $\Psi_\text{gs}$, $\Psi_\text{elastin}$ and $\Psi_\text{collagen}$  represent the free energy contributions from the ground substance, elastin and collagen, respectively.

\begin{figure}[htb!]
    \centering
    \includegraphics[width=0.8\textwidth]{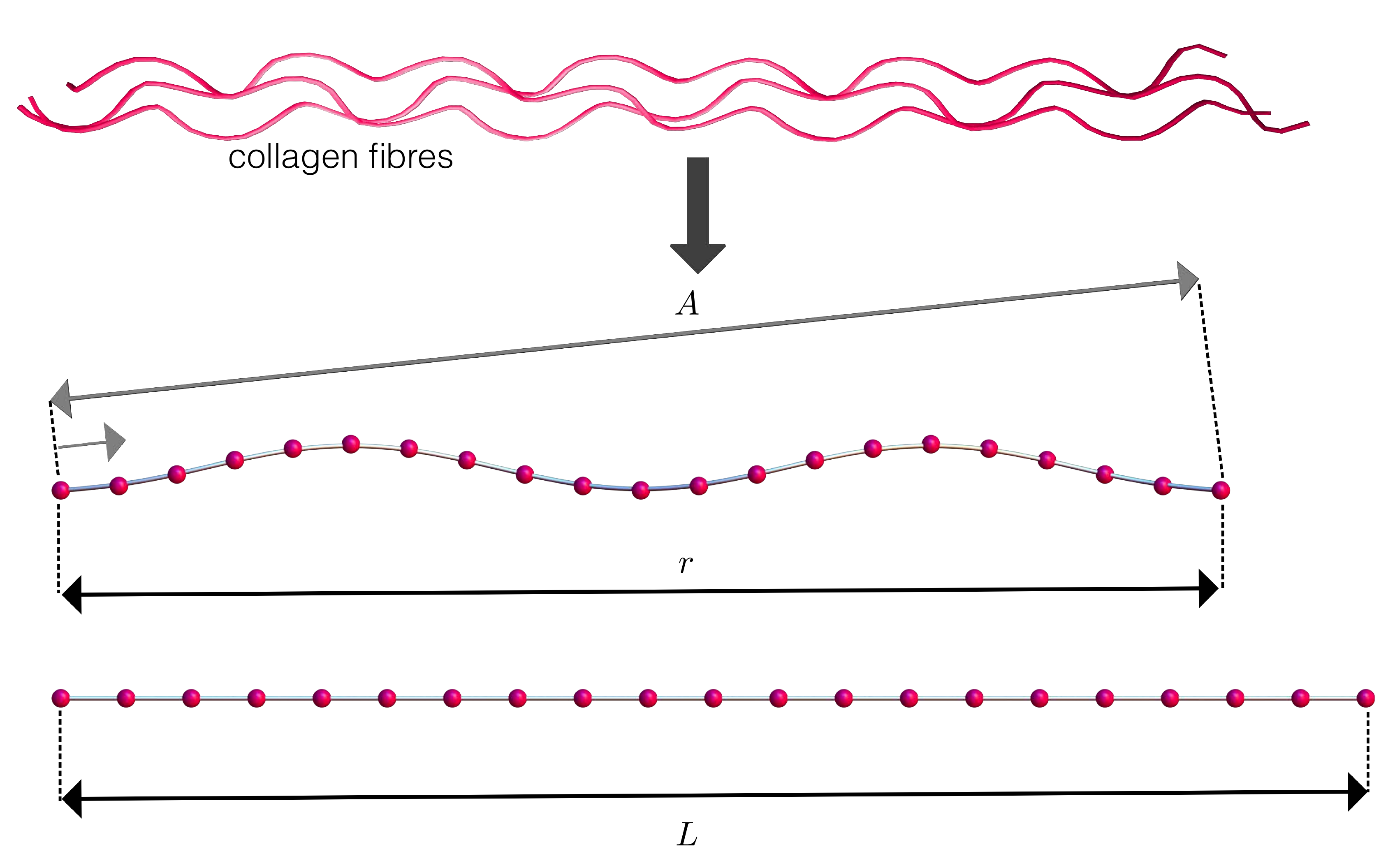}
    \caption{Wormlike chain assembly. The successive chain link is correlated to the chain before it.
    $A$ is the persistence length,  $r$ is the effective end-to-end length and $L$ is the chain contour length.}
    \label{collagen}
\end{figure}

To capture the behaviour at low stretches, the elastin and ground substance energies are given by the following compressible neo-Hookean type free energy
\begin{gather}
  \Psi_\text{gs} + \Psi_\text{elastin} = \left( \alpha_\text{gs} + \alpha_\text{elastin} \right) \left(I_1 - 3 + \dfrac{1}{\beta}\left(I_3^{-\beta}-1\right)\right)
  \label{psi_gs_plus_psi_el}
  \intertext{where}
  \alpha_{(\bullet)} = \dfrac{\mu_{(\bullet)}}{2}, \qquad \qquad \beta = \frac{\nu}{1-2\nu}
\end{gather}
are constitutive parameters, respectively associated with the shear and volumetric response, and $(\bullet)$ is either gs or elastin.
The shear modulus is denoted by $\mu$, and $\nu$ is the ground state Poisson’s ratio of the composite material represented by the ground substance and elastin phases.
Here, the nonlinear, anisotropic and network nature of the collagen response is captured through a transversely isotropic network eight-chain model proposed by \citet{Kuhl2005}.
This model is based on theories related to the micromechanics of macromolecule mechanical networks \citep{Arruda1993, Flory1969} and is a particularisation of the orthotropic eight-chain model developed by \citet{Bischoff2002}.
In these mechanical descriptions of macromolecular networks, long molecular chains are assumed to rearrange their conformation under the influence of random thermal fluctuations (i.e.\ entropic forces).
Unlike polymer chains in rubber \citep{Kuhl2005} which have conformations reminiscent of a random walk, biopolymer chains such as collagen assemblies feature smoothly varying curvature and are therefore considered correlated.
This type of idealised molecular chains is best described using the concept of wormlike chains of \citet{Kratky1949}.
This approach was successfully applied to model the mechanics of DNA molecules by \citet{Marko1995}.
Since then, wormlike chain models have been used to describe the structure and mechanical behaviour of collagen assemblies in the context of skin modelling \citep{Garikipati2004, Bischoff2002, Buganza2012, Flynn2013, Flynn2008, Flynn2009} and other fibrous biological soft tissues \citep{Kuhl2005, Kuhl2007, Bischoff2004, Saez2013}.
It is important to note that, in the development of microstructurally-based constitutive theories, these macromolecular chains could also be defined or interpreted as tropocollagen molecules, collagen micro-fibrils, fibrils, fibres or fibre bundles.
If considering supra-molecular scales, it is clear that the wormlike chain energy is no longer associated with the notion of entropic elasticity and true molecular behaviour but is rather a microstructurally-motivated macroscopic phenomenological  energy that capture well the strain-stiffening behaviour of collagenous structures.
In the present approach, it is assumed that correlated chains represent collagen fibres (\fig{collagen}). This assumption is reasonable and sufficient to demonstrate how microstructural alterations of the dermis as a function of ageing could lead to macroscopic stiffening of the dermis.
In molecular network theories, molecular chains are assumed to be made of beads connected by $N$ rigid links of equal length $d$, the so-called \citeauthor{Kuhn1936} length \citep{Kuhn1936}, so that the maximum length of a chain, the contour length, is $L = N d$.

The mechanics of macromolecular polymer structures is not only governed by the mechanical properties of individual chains but also by their electromagnetic and mechanical interactions which can take the form of covalent bonds, entanglement and physical cross-links.
These combined effects give rise to strong isotropic and anisotropic network properties which can be implicitly and effectively captured by network models such as the eight-chain model of \citet{Arruda1993} or \citet{Kuhl2005}.
The essential assumption underpinning these formulations is that there exists a representative microscopic unit cell able to capture network properties.
The original eight-chain model of \citet{Arruda1993} assumes that the unit cell is made of eight entropic chains of equal lengths connected from the centre of the cell to each of its corners (\fig{collagen}), each equipped with their own entropic energy $\Psi_\text{wormlike}$.
For correlated chains, one would consider the following wormlike chain energy defined by
\begin{equation} \label{psi_wormlike}
  \Psi_\text{wormlike} = \Psi_0 + \dfrac{k\theta L}{4A}\left(2\dfrac{r^2}{L^2} + \dfrac{1}{(1-\frac{r}{L})} - \dfrac{r}{L}\right) \, ,
\end{equation}
where $L$, $A$, $r_0$ and $r$  and  are respectively the contour, persistence, initial end-to-end length and the current end-to-end length of the chain (see \fig{collagen}), and $\Psi_0$ is the wormlike chain energy in the unperturbed state.
The wormlike chain has the defining characteristic that the chain segments are correlated and exhibit a smooth curvature along the contour.
This correlated form is captured by the persistence length $A$.
The persistence length can be viewed as a measure of stiffness.
\citet{Garikipati2004} refers to it as a measurement of the degree to which a chain departs from a straight line, while \citet{Marko1995} interpret it as the characteristic length over which a bend can be made with energy cost $k \theta$, where $k = \SI{1.38064852}{\joule\per\kelvin}$ is the Boltzmann constant and $\theta$ the absolute temperature.

\begin{figure}[htb!]
    \centering
    \includegraphics[width=\textwidth]{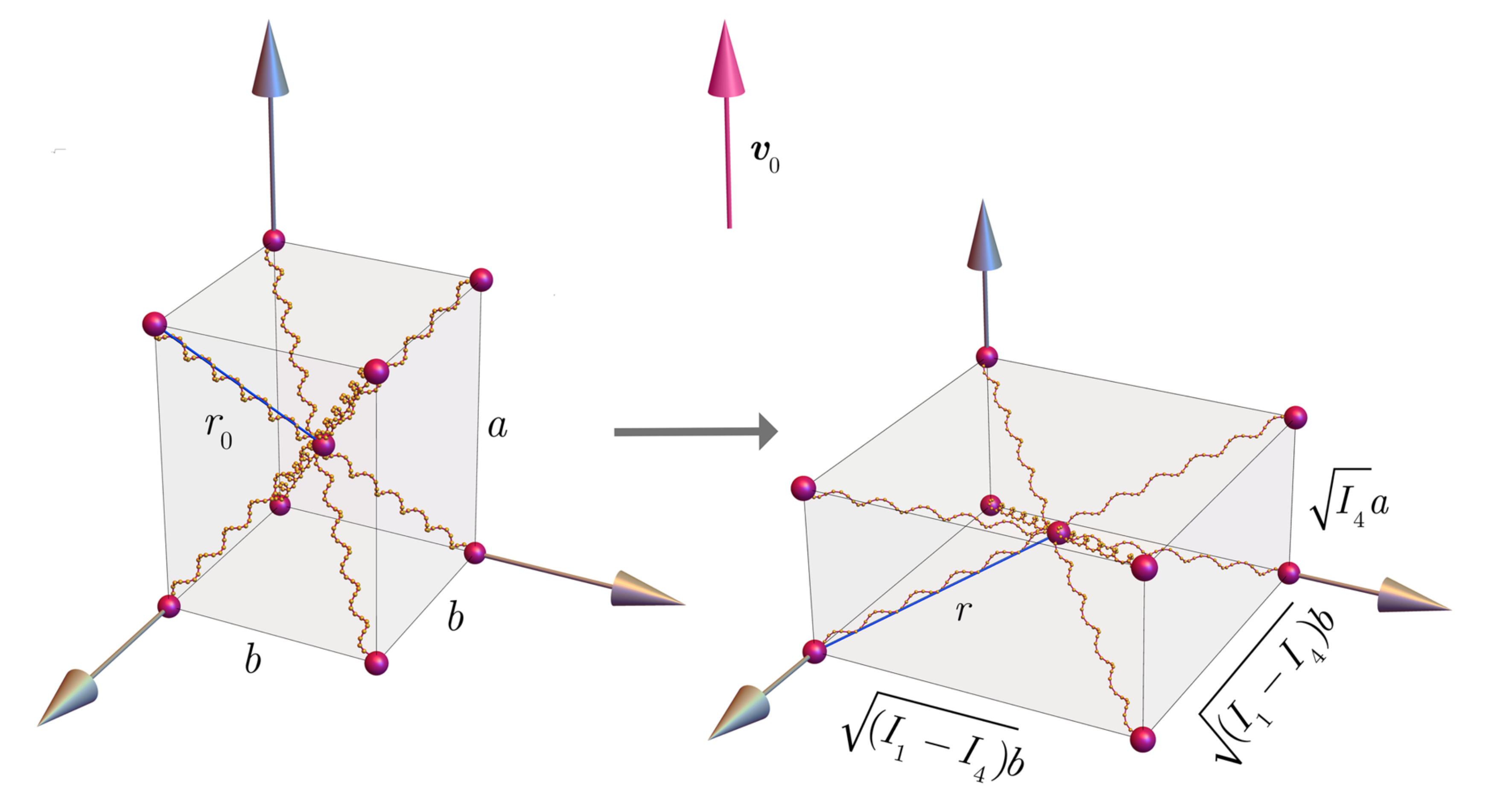}
    \caption{Schematic representation of the transversely isotropic eight chain network model of \citep{Kuhl2003}.
    The eight polymer chains are governed by a wormlike energy function $\psi_\text{wormlike}$ (see \eqn{psi_wormlike}).
    The unit cell dimensions are $a$ and $b$ while unit vector $\b{v}_0$, aligned with one of the principal direction of the unit cell, is the unit vector corresponding to the preferred orientation of the collagen network in the undeformed configuration.}
    \label{8_chain}
\end{figure}

In order to incorporate such chain models into an invariant-based constitutive framework it is necessary to relate the individual chain stretch to the overall or macroscopic deformation.
To that end, the principle of affinity is invoked so that the macroscopic and unit cell principal directions are identical.
Due to the symmetry of the chain structure, the stretch of each chain can be found as a function of the principle stretches.
\changemarker{In \fig{8_chain} a unit cell arrangement of dimensions $a \times b \times b$ is depicted.}
 For the case of anisotropy $a \neq b$.
Additionally, the unit cell is characterised by the unit vector $\b{v}_0$  that corresponds to the (local) preferred orientation of the collagen network. The undeformed end-to-end length of each of the individual chains is given by
\begin{equation*}
  r_0  = \dfrac{1}{2} \sqrt{a^2 + 2b^2} \, .
\end{equation*}
The deformed end-to-end length $r$ can be expressed using the invariants defined in \eqn{I1_to_I3} and \eqn{I4} by
\begin{equation*}
  r  = \dfrac{1}{2} \sqrt{I_4 a^2 + (I_1 - I_4)b^2} \, .
\end{equation*}

The invariants $I_1$, $I_3$ and $I_4$  are characteristic of the macroscopic deformations of the polymer.
It is assumed that the microscopic stretch along the  direction $\b{v}_0$ is captured by the macroscopic term $I_4$.
This effectively couples the microscopic and macroscopic length scales giving rise to a multiscale model. The relative stretch of a collagen fibre is defined by $\lambda_\text{r} := r / r_0$.

The final term of \eqn{psi} is further decomposed into the additive form as
\begin{equation*}
  \Psi_\text{collagen} = \Psi_\text{chain} + \Psi_\text{repulsive} \, ,
\end{equation*}
where $\Psi_\text{chain}$ reflects the effective assembly of the eight chain energies, i.e.\  $\Psi_{\text{chain}} := \gamma_{\text{chain}} \Psi_{\text{wormlike}}$ and $\Psi_\text{repulsive}$ is a repulsive energy that ensures the initial configuration is stress free and that the chain does not collapse.
$\Psi_\text{chain}$ denotes the chain density per unit cell.
\citet{Saez2013} interprets $\gamma_\text{chain}$ as the number of molecules within a collagen fibril, while \citet{Bischoff2002} describes it as the collagen fibre density which in the context of collagen, this corresponds to the number of fibres within a bundle.
It follows that
\begin{gather*}
  \Psi_{\text{chain}} = \dfrac{\gamma_{\text{chain}}kL\theta}{4A}\left(2\dfrac{r^2}{L^2} + \dfrac{1}{1-\frac{r}{L}} - \dfrac{r}{L}\right) \, , \\
  \Psi_{\text{repulsion}} = -\dfrac{\gamma_{\text{chain}}k\theta}{4A}\left(\dfrac{1}{L} + \dfrac{1}{4r_0}\dfrac{1}{1-\frac{r_0}{L}}- \dfrac{1}{4r_0}\right)\left(\ln\left(I_4^{[a^2-b^2]/2}\right)+ \dfrac{3}{2}\ln\left(I_1^{b^2}\right)\right) \, .
\end{gather*}
All length quantities have been normalised  by the link length $L$ which is the length of a single link along the wormlike chain assembly.
A description of the constitutive parameters of $\Psi_\text{collagen}$ is given in \tableref{table_psi_collagen_parameters}.

\changemarker{The expression for the first Piola--Kirchhoff stress $\b{P}$ then follows directly from \eqn{PK2}.}

\begin{table}[htb!]
\centering
\begin{tabular}{ l l l }
 \toprule
\textbf{Parameter} &  \textbf{Symbol}   &  \textbf{Units} \\
 \hline
 Poisson's ratio & $\nu$  &  -\\
 Shear modulus & $\mu$ &  \si{\newton\per\square\metre}\\
 Boltzmann constant & $k$  &  \si{\joule\per\kelvin}\\
 Absolute temperature & $\theta$ &  \si{\kelvin} \\
 Chain density & $\gamma_\text{chain}$  & \si{\metre\cubed}  \\
 Contour length & $L$  &  - \\
 Persistence length & $A$ & -\\
 Unit cell dimensions & $a,b$ & -
  \\
\bottomrule
\end{tabular}
\caption{Constiutive parameters of $\Psi_\text{collagen}$.}
\label{table_psi_collagen_parameters}
\end{table}


\subsection{Weak form of the governing relations}

The weak form of the governing and the accompanying Neumann boundary condition \changemarker{in \eqns{balance_momentum}{neumann_bc}} is essential for establishing the approximate solution using the finite element method (FEM).
\changemarker{Following the Principle of Virtual Working, multiplying \eqn{balance_momentum} by an vector-valued test function $\delta \b{u}$, where $\delta \b{u} = \b{0}$ on $\partial \Omega_{0,\text{D}}$}, and integrating by parts the result over the reference configuration $\Omega_0$ yields the expression of the residual for the \changemarker{equilibrium equation as}
\begin{align}\label{weak_form}
  \int_{\Omega_0} \Grad \delta \b{u} : \b{P} \, \d V
  - \int_{\partial \Omega_{0,\text{N}}} \delta \b{u} \cdot \o{\b{T}} \, \d A  = 0 \, .
\end{align}

\section{Microstructurally-based mechanistic modelling of ageing}\label{sec_micro_modelling_age}

The central hypothesis underpinning this study is that the effects of ageing on mechanical properties of the tissue are directly linked to alterations in the microstructural characteristics of the collagen and elastin networks.
In this section, some of the constitutive parameters introduced in \sect{sec_continuum_model_skin} are motivated as plausible mechanistic descriptions of the intrinsic ageing process as evidenced by experimental observations (see \tableref{table_const_param_lik_ageing}).
It is worthy to point out that, in reality, the material and structural effects of ageing are likely to be coupled and would therefore lead to a dependency between constitutive parameters.
For sake of simplicity, and in the absence of relevant experimental data, this interplay is not accounted for here.
The effect of intrinsic ageing on the ground substance manifests itself as a more pronounced viscoelastic response with age.
As the constitutive model for the ground substance only capture the elastic response it is assumed that ageing does not affect the ground substance related parameters in \eqn{psi_gs_plus_psi_el}.

\begin{table}[ht!]
\centering
\begin{tabularx}{\textwidth}{ l l l  }
 \toprule
 \bf{Constituent} &  \bf{Ageing effect} & \bf{Parameter} \\
 \hline
  Ground substance & \parbox{7cm}{Increased water content results in more pronounced viscoelasticity} & n/a \\ \midrule
  Elastin & Loss of elasticity & $\alpha_\text{elastin}$ \\ 
  & Destruction of elastin network & $\alpha_\text{elastin}$ \\ \midrule 
  Collagen & Loss of mature collagen & $\gamma_\text{chain}$\\ 
  & Increased crosslinking  & $\gamma_\text{chain}$ \\
  & \parbox{7cm}{Flattening and unravelling of collagen network} & $a,b$ \\ 
  & Alterations in anisotropy & $a,b$\\
\bottomrule
\end{tabularx}
\caption{Description of the constitutive parameters the evolution of which is associated with ageing.}
\label{table_const_param_lik_ageing}
\end{table}

The elastin network is observed to degrade with age.
The loss of elastic integrity would logically lead to a reduced contribution from the elastin component to the composite strain energy function (\eqn{psi}) of the constitutive model.
One approach could have been to introduce a volume fraction for elastin, ground substance and collagen.
Degradation of the elastin network would have then been modelled as a reduction in volume fraction of elastin.
Here, instead of using the latter approach, the value of $\alpha_\text{elastin}$ is assumed to decrease with age.
It should be noted again that elastin is a highly-stable protein which undergoes little turnover before the age of 40.
Thus $\alpha_\text{elastin}$ would remain constant for the first four decades.

Within the current model, there is no natural parameter that explicitly describes the level of collagen crosslinking at any point in time.
It is believed that the increased crosslinking with age is one factor responsible for the increased stiffness of the skin. Through increasing $\gamma_\text{chain}$, it is expected that a stiffer response will be elicited.
Although $\gamma_\text{chain}$ describes the number of fibres within a bundle, it represents the closest link between the observed increase in crosslinking with age and the model.
It would be expected that a decrease in collagen density would contribute to a decrease in the overall stiffness of the skin as the collagen network loses integrity.
Thus it is reasonable to assume that a decrease in $\gamma_\text{chain}$ with age could describe a loss of collagen.
It is impossible to decode the complex interplay between these two factors.

As discussed previously, the slope of the linear region of the stress-strain curve for skin, as a composite structure, remains constant with age but the onset of strain hardening occurs for lower strains.
This suggests that there might be a mechanism that simply alters the structural characteristics of the collagen network and not the mechanical properties of collagen microfibrils.
In their unloaded states collagen fibres are crimped.
It is believed that this conformational state is due to the presence of highly cross-linked elastic fibres — mainly elastin fibres \citep{Kielty2002} - and is also the result of active tensions exerted by fibroblasts.
Any reduction in fibroblast density in the collagen dermal network would have an effect on these active forces and on the rest state of collagen fibres.
Such a reduction would tend to relax the residual strain/tension in the dermis.
This would make collagen fibres less crimped and therefore their apparent length would be closer to their fully taut length.
Thus, when the skin is macroscopically loaded in tension, the dermal tissue will not exhibit a very pronounced toe region — corresponding to the progressive uncrimping of collagen fibres — but will rather reach the stiffer linear region much quicker.
The corollary of this observation is that for a given applied macroscopic strain to the skin, the response of the intrinsically-aged dermal collagen network will be stiffer.
With age, due to dermal flattening and the loss of elastic recoil, the collagen fibres are observed to uncrimp \citep{Sherratt2013}.
In the chain network structure of the model, this is captured by adjustments to the end-to-end length of the collagen fibres through the parameters $a$ and $b$.
By increasing $a$ and $b$ there will be a reduction in the range of strain of the low modulus portion prior to the onset of the collagen response at lower stretches.
Thus by increasing $a$ and $b$ the model should be capable of capturing the stiffening response. Additionally, the anisotropic response of the skin is observed to change with age.
\citet{Tonge2013b} reports a loss of anisotropy through observations of the bulge test.
As the model developed here will be compared to the bulge tests at various ages, it is this behaviour that is intended to be captured.
By reducing the ratio of $a / b$ , either by increasing $b$ or decreasing $a$ or a combination of both, this should be obtained.
Undeniably, there is a reduction in the thickness of the skin with age.
Such an observation is backed by numerous sources in the literature \citep{Diridollou2001, Escoffier1989, PaillerMattei2008}, as mentioned in the previous section, but there is still much debate over the exact relationship between age and skin thinning.
The dermal thickness changes with age which will be captured directly in the geometry.

\section{Numerical simulations of the bulge test}\label{sec_FE_bulge_test}

\subsection{Details of the finite element implementation}

The highly-nonlinear system of governing relations \eqref{weak_form} are solved using the finite element method.
The system of equations is linearised using a global Newton--Raphson procedure.
\changemarker{The finite element library AceGen / AceFEM \citep{Korelc2002, Korelc2016} is used to implement the solution scheme.
AceGen facilitates the generation of the finite element interpolation and constitutive model by exploiting symbolic and automatic differentiation, and automated code generation.
AceFEM provides a flexible environment for the iterative solution of the nonlinear finite element problem.}
An adaptive incremental loading scheme is employed to ensure convergence of the numericlal scheme.

\subsection{Parameter identification}

In the literature, multiple mechanical skin tests have been proposed to characterise the skin response to loading.
The variety of skin tests and  different methods to perform them, as well as the natural skin variation that exists through factors such as ethnicity, gender, age and anatomical site has resulted in a broad and varying characterisation of skin behaviour.
In general, \emph{in vivo}, in both physiological and supra-physiological conditions, the skin exhibits material nonlinearity, anisotropy, viscoelasticity and near incompressibility and can also sustain large deformations \citep{Limbert2017}.
However, elicitation and relevance of these characteristics is highly variable.
Although there are many experimental techniques to characterise certain aspects of skin mechanics \citep{Jor2013}, bulge tests are considered here.
Performed \emph{in vitro}, the bulge test applies a positive pressure to the underside of an excised skin sample.
The skin sample is fixed at a specified diameter aperture which allows for ``bulge-like'' deformation of the sample under pressure.
Deformations are measured and linked to the state of stress so that constitutive parameters can be identified \citep{Tonge2013, Tonge2013b}.
The choice of the bulge experiment as a test-bed for the constitutive and computational and models developed here is motivated by the following factors:
\begin{itemize}
  \item As the experiments are performed \emph{in vitro}, the boundary conditions of the numerical model are simpler to define and impose.
  \item The large deformations induced in the physical tests ensure that the various components of the constitutive model would contribute.
  At small deformations the elastin and ground substance will assume the dominant role, and ultimately, as macroscopic strains increase, the collagen fibres will become active and dictate the majority of the response at large deformations.
  \item The specimens were taken from the back of the patients, thus minimising any photoageing effects such as those observed on sun-exposed regions (e.g.\ the face).
  \item The experimental results published by \citet{Tonge2013, Tonge2013b} provided a broad range of kinematic measurements with which to compare the models.
  \item Crucially, for the present study, the experiments by \citet{Tonge2013, Tonge2013b} were conducted on skin samples harvested from donors featuring a broad range of ages.
  This allows the correlation between age and mechanical properties (i.e.\ constitutive parameters) to be studied.
\end{itemize}

The procedure for the bulge test follows that detailed in \citep{Tonge2013} where $10 \times 10$ \si{\cm}  skin specimens were procured from the back torso of donors, ages ranging from 43 to 83 years.
After excision, the adipose tissue was removed and tissue thicknesses measured.
The thickness, gender, age and anatomical site of the samples are listed in \tableref{table_tonge_data}.
Thickness was measured at the centre of each edge and an average taken.
This average was used as the uniform thickness of the sample in the numerical model.

\begin{table}[h!]
\centering
\begin{tabular}{ l l l l }
\toprule
 \bf{Age}  &  \bf{Gender} & \bf{Site} & \bf{Thickness} [\si{\mm}] \\ [0.5ex]
\hline
 43 & Male & Lower back & 4.86 \\
 44 & Male & Lower back & 4.38 \\
 59 & Female & Unknown & 5.18 \\
 61 & Male & Left upper back & 2.01\\
 62 & Female & Unknown & 2.95 \\
 83 & Male & Unknown & 2.43 \\
\bottomrule
\end{tabular}
\caption{Donor and specimen information from \citet{Tonge2013}.}
\label{table_tonge_data}
\end{table}

\begin{figure}[htb!]
    \centering
    \includegraphics[width=0.4\textwidth]{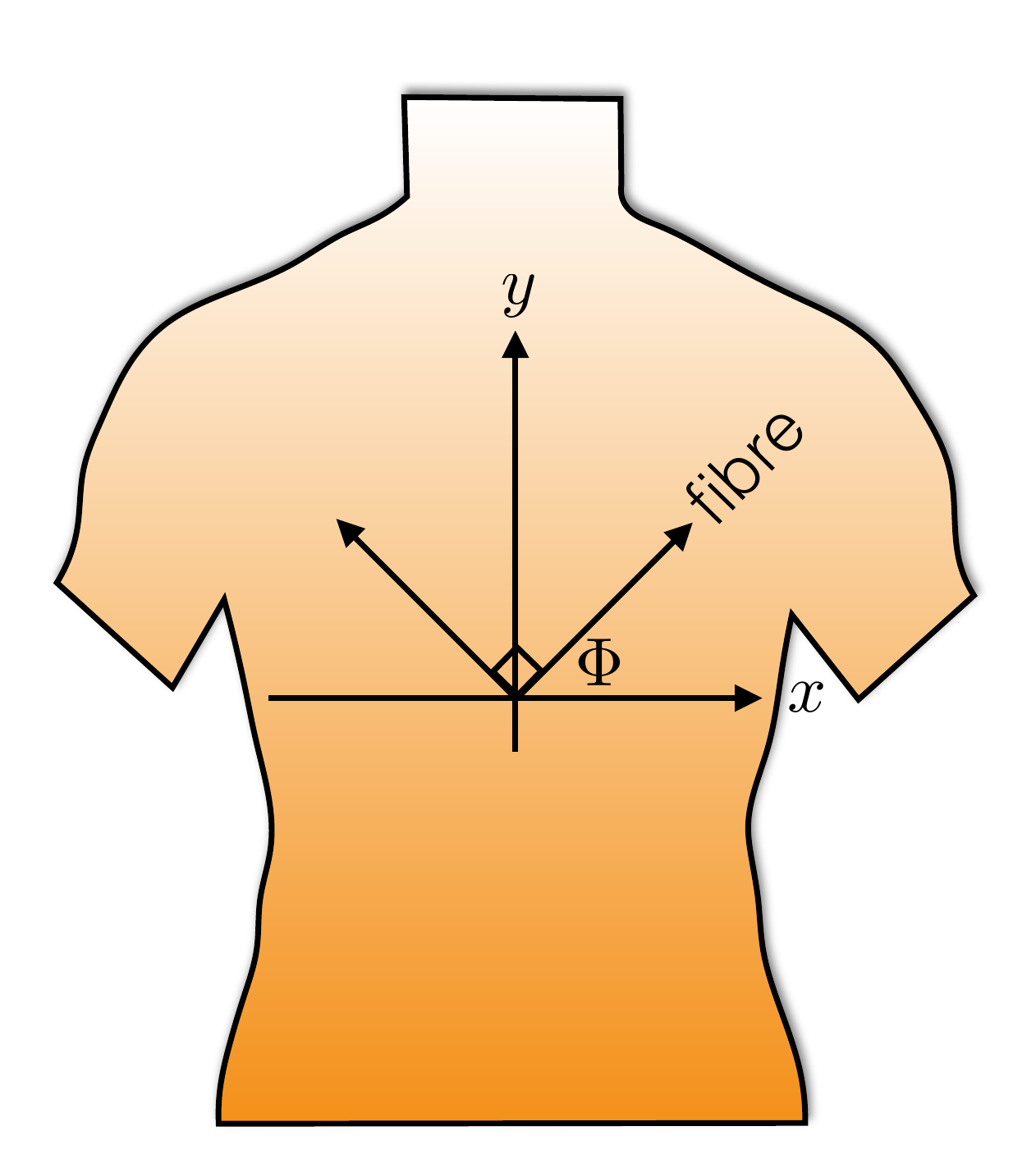}
    \caption{Body axes: the $y$-axis corresponds to the vertical body axis and the $x$-axis to the horizontal body axis.
    Recreated from \citet{Tonge2013b}.}
    \label{skin_sample_location}
\end{figure}

The specimens were glued to a \SI{7.5}{\cm} diameter ring.
\changemarker{The ring serves to constrain the skin specimen.
The skin interior to the ring is subject to pressure loading while that exterior is fully constrained.}
The coordinate system for the samples was set such that the $y$-axis corresponded to the vertical body axis and the $x$-axis to the horizontal body axis, as shown in \fig{skin_sample_location}.
Fibre and perpendicular directions can thus be defined by the angle $\Phi$ from the horizontal axis.
Controlling for relative humidity and temperature, samples were inflated through a controlled applied pressure, with a maximum pressure of \SI{5.516}{\kPa}.
Upon inflation the skin samples deformed to an elliptical dome.
The dimensions of the ellipse are used to determine the dominant fibre direction with $\Phi$ defined accordingly.

\begin{figure}[htb!]
    \centering
    \includegraphics[width=0.8\textwidth]{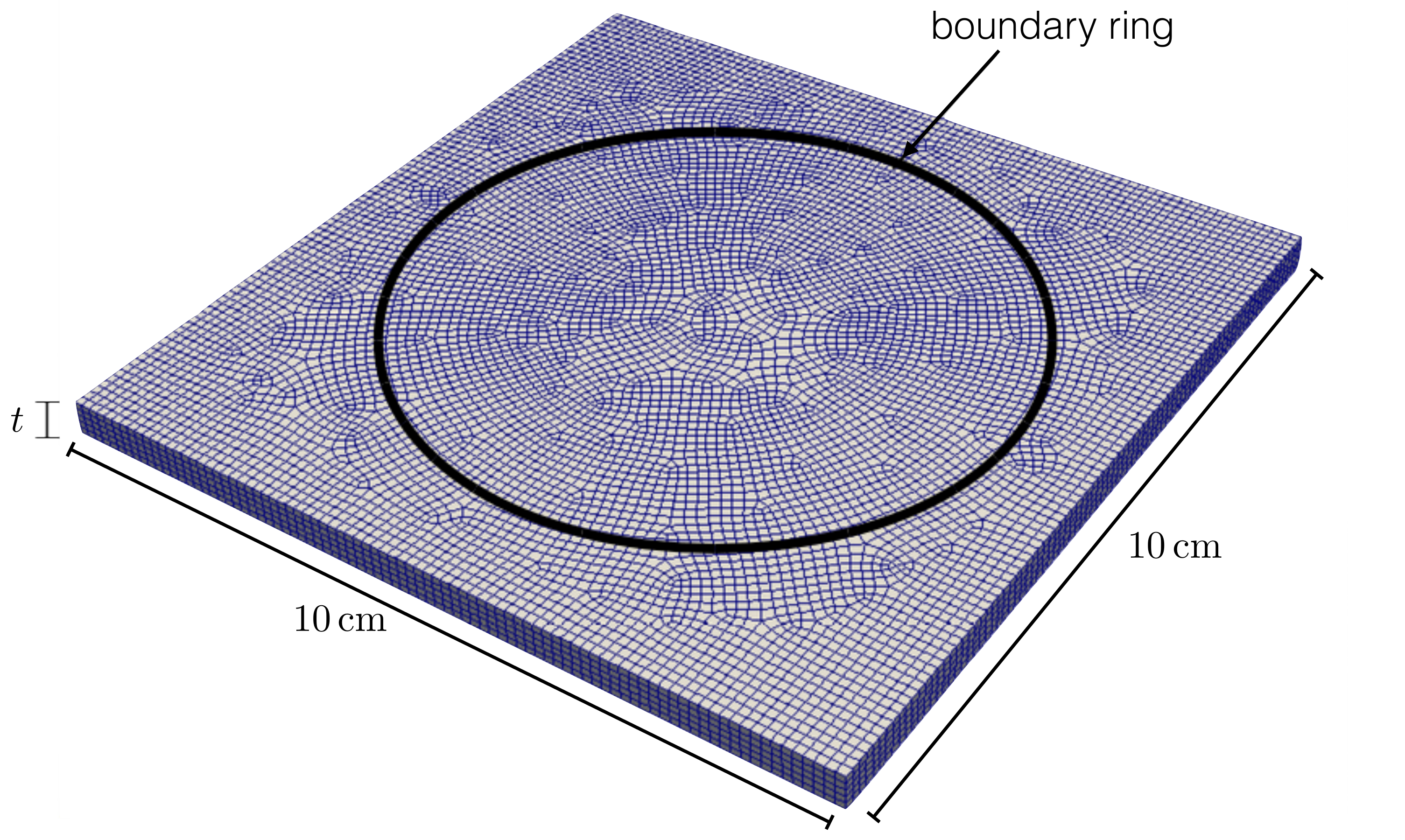}
    \caption{Finite element mesh used for the bulge tests.}
    \label{fe_mesh}
\end{figure}

In this part of the study, the objective is to use a finite element updating technique \citep[see e.g.][]{Jor2011, Jor2013, Kvistedal2009} to identify the constitutive parameters of the transversely isotropic eight-chain model of the skin for various ages.
\fig{fe_mesh} shows the discretisation of the skin sample.
The skin thickness is set to the dermal thicknesses as specified in \tableref{table_tonge_data}.
The geometry is discretised using 8-noded hexahedral trilinear elements, with 5 elements through the thickness.
Through a mesh sensitivity study, it was found that \num{46416} nodes ensured a sufficiently converged solution.
The nodes on the upper surface outside the bulge diameter of \SI{7.5}{\cm} and the outer edges are held fixed.
The pressure loading condition is applied to the bottom surface.
As proposed in the previous Section, only the original network dimensions $a$ and $b$, $\gamma_\text{chain}$ and $\alpha_\text{elastin}$ are assumed to be variable.
The rest of the parameters are assumed age invariant and are given by  $L=2.125$, $A=1.82$, $\gamma_\text{gs} =\SI{100}{\Pa}$, $\beta = 4.5$, $\theta = \SI{310}{\kelvin}$.
Furthermore, only the tests for male specimens are simulated in order to exclude the influence of variations between genders.
\changemarker{The age-dependent parameters that gave the best fit to the published bulge test-data were identified using an iterative manual process.}

\begin{figure}[htb!]
    \centering
    \includegraphics[width=\textwidth]{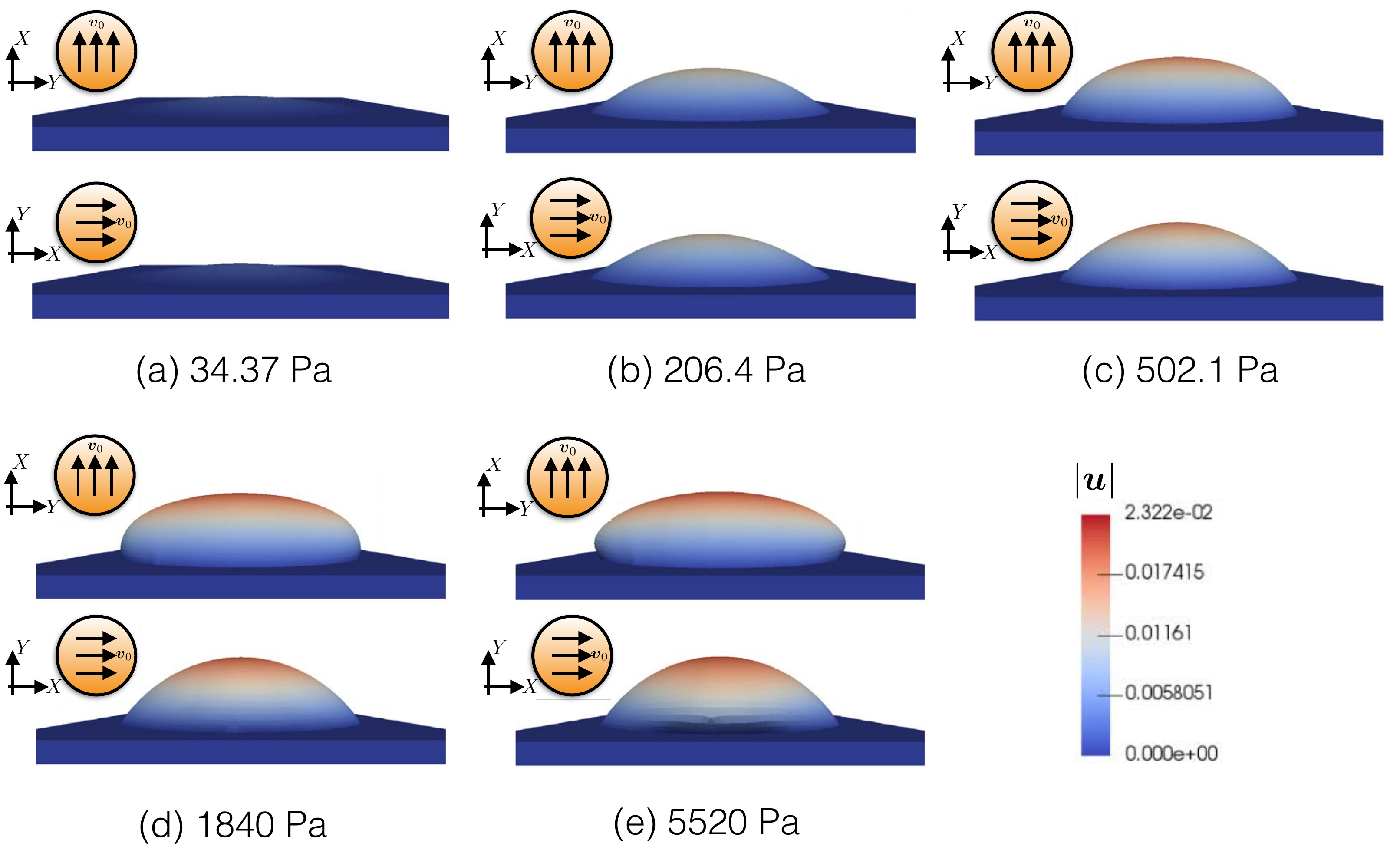}
    \caption{Displacement profiles for the 44 years-old male skin specimen: (a) pressure = \SI{34.37}{\Pa}; (b) pressure = \SI{206.4}{\Pa}; (c) pressure = \SI{502.1}{\Pa}; (d) pressure = \SI{1840}{\Pa}; (e) pressure = \SI{5520}{\Pa}.}
    \label{bulge_deformed_shapes}
\end{figure}

In \fig{bulge_deformed_shapes}, the profile of the skin obtained from the finite element simulation of the bulge test on the 44 years-old male specimen is given at monotonously-increasing pressure loading from \SI{34.37}{Pa} to \SI{5.52}{\kilo\Pa}.
In each figure, the profiles are given along and perpendicular to the fibre direction $\b{v}_0$ as illustrated by the arrows in each subfigure.

\begin{figure}[htb!]
    \centering
    \includegraphics[width=\textwidth]{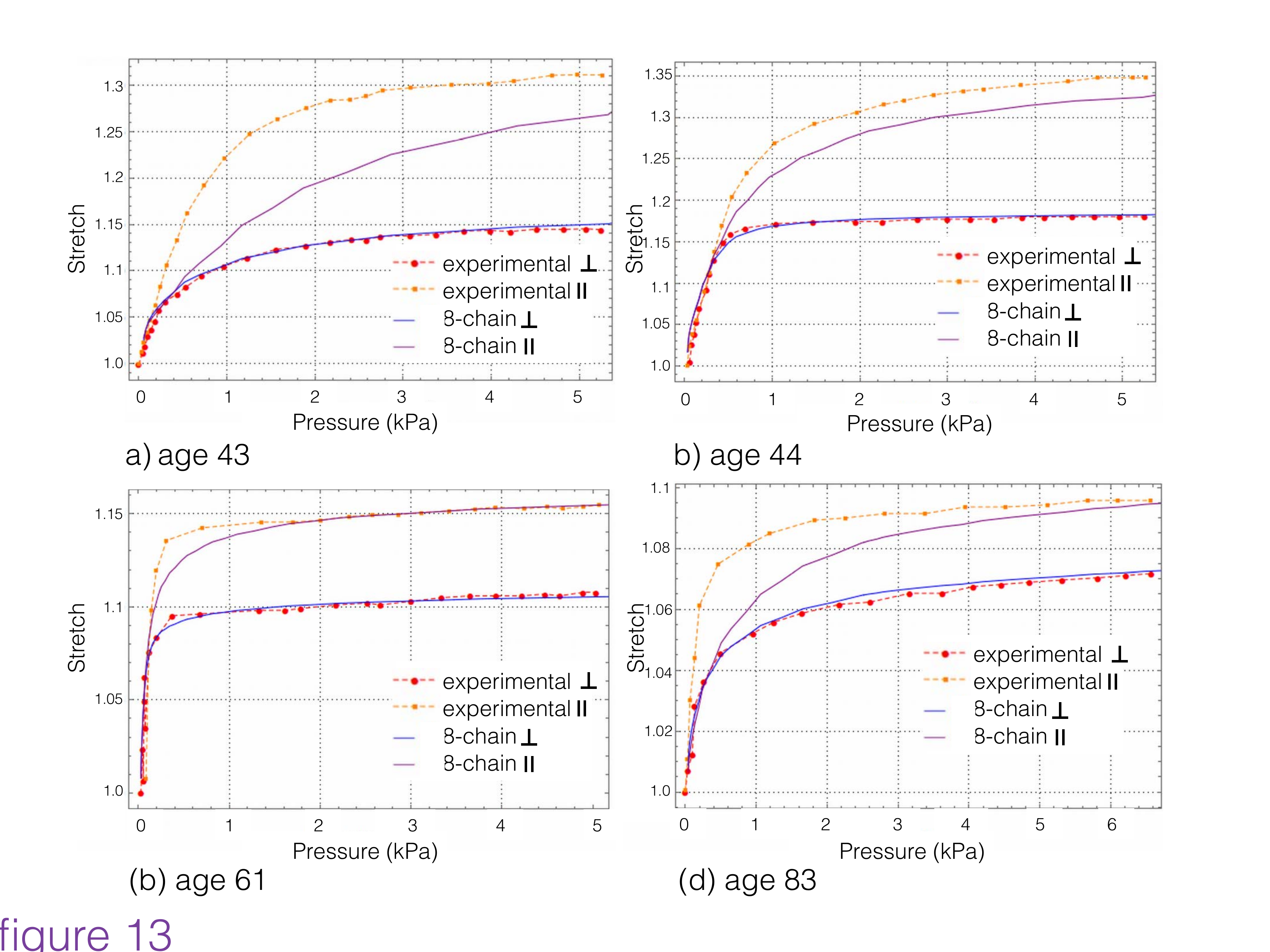}
    \caption{Plots of in-plane stretch as a function of inflation pressure for the experimental measurements of \citet{Tonge2013b} overlaid over the theoretical curves obtained from the identification of constitutive parameters.}
    \label{stretch_pressure}
\end{figure}

The bulge specimen undergoes a rapid initial displacement while the stress state is still within the low modulus portion of the stress-strain curve.
This can be seen in the plot of the pressure versus the in-plane stretch  where a small increment in pressure induces a large change in stretch (\fig{stretch_pressure}).
During this phase, the elastin and ground substance contributions dominate the mechanical response and offer little resistance to inflation.
The divergence of the pressure-stretch lines indicates the activation of the anisotropic collagen network.
In terms of the material model, the collagen energy contribution is no longer negligible as the end-to-end chain length $r$ sustains significant stretch and approaches the contour length $L$.
For the age 44 specimen, above \SI{400}{\Pa} further stretch parallel to the fibre direction no longer occurs whereas perpendicular stretch continues to increase with pressure.
This results in the response shown in \fig{bulge_deformed_shapes}(c) where at \SI{502}{\Pa} the profiles along the perpendicular and parallel directions differ.
Mechanical anisotropy is amplified with increased pressure as the perpendicular stretch increases. As the specimen is deformed, exponentially more pressure is needed to attain further deformation which is characterised by the locking type limit for many biological soft tissues and engineered polymers.
In \fig{bulge_deformed_shapes}(d) and \fig{bulge_deformed_shapes}(e) displacement at high pressure occurs laterally in the perpendicular fibre direction.

The 8-chain model is now used to simulate the stretch behaviour of the skin at increasing pressure at increasing age, and the results summarised in \fig{stretch_pressure}.
The stretch is measured parallel and perpendicular to the dominant fibre direction.
In general, the stretch behaviour with increasing age is reasonably captured.
The fit at age 43 is notably poor which suggests that there is an additional directional dependence not captured by the model or a missing constituent contribution (i.e.\ an incomplete constitutive model), or experimental error.

Excluding the response of the age 43 test, the identified parameters shown in \tableref{skin_identified_parameters} are in agreement with the hypotheses advocated in \sect{sec_struct_mat_props}.
Note, $\gamma_\text{chain}$ does not vary monotonically with age, with the same values obtained for age 44 and 83.
This suggests that $\gamma_\text{chain}$ is not age dependent.
The difference in the initial end-to-end lengths as dictated by the increased values of $a$ and $b$ allows for the difference in maximum stretches observed in the experiments.
Additionally the ratio between the two changes with age with $a : b =  4.313 : 1, 2.84 : 1, 2.836 : 1$ at age 44, 61 and 83 respectively.
This reduction, especially between age 44 and 61 is indicative of the expected reduction in anisotropy.
The value of $\alpha_\text{elastin}$ decreases almost linearly with age, with an approximate reduction of \SI{100}{\newton\per\meter\squared} every five years.

\begin{center}
\begin{table}[ht!]
\centering
\begin{tabular}{l l l l l}
 \toprule
 \bf{Parameter} & \bf{Age 43} &  \bf{Age 44} &   \bf{Age 61} &   \bf{Age 83} \\
 \hline
 $a$ & 3.58 & 3.45 & 3.55 & 3.65\\
 $b$ & 0.5 & 0.8 & 1.25 & 1.287\\
 $\gamma_\text{chain}(\times 10^{22})$ & $6$ $\text{m}^{-3}$  & $0.856$ $\text{m}^{-3}$ &  $0.0856$ $\text{m}^{-3}$ & $0.856$ $\text{m}^{-3}$ \\
 $\alpha_\text{elastin}$ & 1000 \si{\newton\per\square\metre} & 1300 \si{\newton\per\square\metre} & 1000 \si{\newton\per\square\metre} & 500 \si{\newton\per\square\metre}\\
 \bottomrule
\end{tabular}
\caption{Wormlike 8-chain model parameter values for the age 43, 44, 61 and 83 bulge test.}
\label{skin_identified_parameters}
\end{table}
\end{center}

\section{\changemarker{Modelling of ageing}} \label{sec_validation_age_model}

\begin{figure}[htb!]
     \centering
     \includegraphics[width=\textwidth]{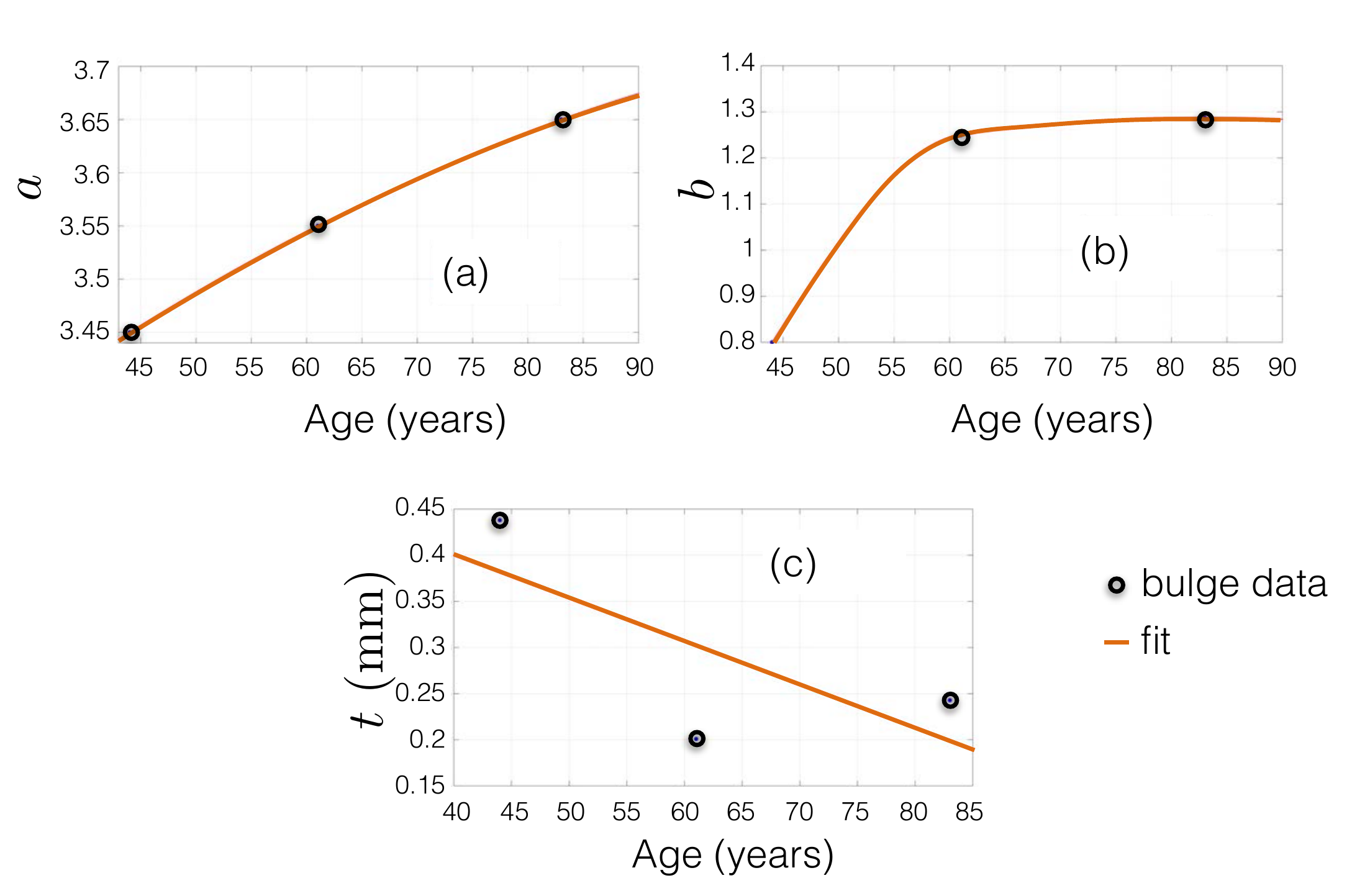}
     \caption{Network dimension and skin thickness trends with age as found through bulge test fits.}
     \label{t_vs_age}
 \end{figure}

The factors discussed in the previous sections provide a means by which the skin constitutive model (and its associated finite element model replicating bulge tests) can capture ageing effects in a continuous sense through the modification of a few selected parameters.
Using the $a$ and $b$ parameter values as found from the age 44, 61 and 83 fits, general ageing trends were established.
It is clear that a mere three data points is insufficient in order to conclusively determine an ageing trend, but they are adequate for the current proof of concept.
Shape-preserving fitting algorithms in Matlab (The MathWorks Inc., Natick, MA, USA) were used to find a continuous evolution of $a$ and $b$ as shown in \fig{t_vs_age}(a) and \fig{t_vs_age}(b).
As expected, there is a general increase in both parameters with age.
$a$ evolves almost linearly with age whereas $b$ undergoes a large increase between age 44 and 61 followed by a plateau.
 This suggests that between age 44 and 61 there is a more significant loss of anisotropy due to a realignment or redistribution of the collagen fibres as a network as compared to later in life.

The skin thickness data as presented in \tableref{table_tonge_data} was used to determine a linear equation for the skin thickness as a function of age. From \fig{t_vs_age}(c) it can observed that a very rough fit has been established due to highly irregular data.
The linear fit is sufficient and is given by
\begin{equation} \label{t_fit_with_age}
  t = -0.0047[\text{mm/year}] \times \text{age}\,[\text{years}] + \SI{0.59}{[\mm]} \, .
\end{equation}

The overall decrease in skin thickness is consistent with experimental values found in the literature.
With reference to \eqn{t_with_time}, the gradient of the proposed relation is comparable.
The difference in the $y$-intercept value is due to \eqn{t_with_time} accounting for thinning from the age of 60 while \eqn{t_fit_with_age} does not account for the fact that thinning does not occur earlier in life.

With the established trends, bulge test simulations were run at 5 year increments.
\tableref{table_parameters_with_age} contains the adjustments to the parameters of interest.
As mentioned, $\alpha_\text{elastin}$ decreases from \SI{1300}{\N\per\m^2} at age 44 by \SI{100}{\N\per\m^2} every 5 years.

\begin{table}[htb!]
\centering
\begin{tabular}{l*{3}{c}}
\toprule
   Age   &  $a$  &  $b$  &  $\alpha_\text{elastin}$  \\ \hline
    49  & 3.483 & 0.984 & 1200  \\
    54 & 3.513 & 1.14 & 1100  \\
    59 & 3.54 & 1.236 & 1000  \\
   64 & 3.564 & 1.26 & 900  \\
   69 & 3.587 & 1.273 & 800  \\
   74 & 3.613 & 1.282 & 700  \\
   79 & 3.634 & 1.287 & 600  \\
   84 & 3.654 & 1.288 & 500  \\
   89 & 3.671 & 1.285 & 400  \\
  \bottomrule
\end{tabular}
\caption{Modification of age dependent model parameters.}
\label{table_parameters_with_age}
\end{table}

\begin{figure}[htb!]
    \centering
    \includegraphics[width=\textwidth]{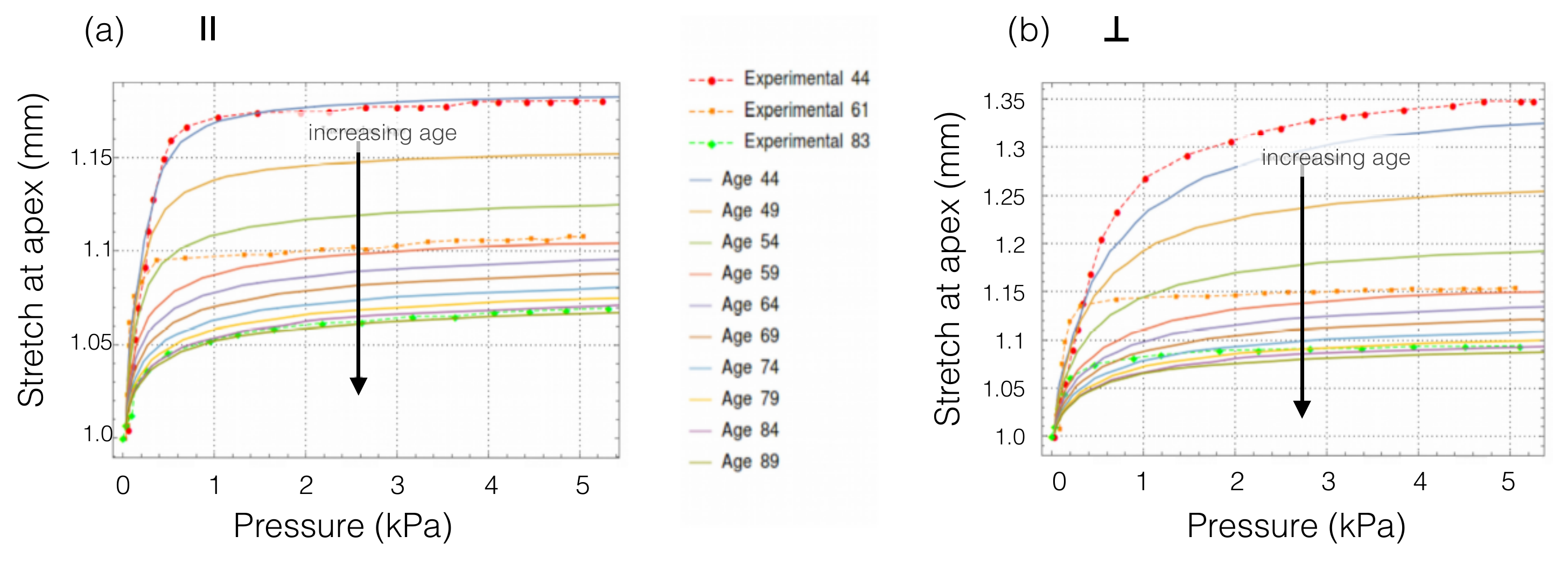}
    \caption{Stretch at the apex of the pressurised skin as a function of applied pressure for 9 age values (see \tableref{table_parameters_with_age}); (a) parallel to the fibre direction; (b) normal to the fibre direction.}
    \label{p_vs_stretch_with_age}
\end{figure}

In \fig{p_vs_stretch_with_age} the simulated evolution of the stretch behaviour with age is given, and compared to the experimental data at ages 44, 61 and 83. As expected, the general ageing-induced trend is captured as the skin stiffens with age.
The maximum stretch obtained drops significantly between ages 44 and 59, with less significant drops thereafter.
This is to be expected primarily due to the trend in parameter b as given in \fig{p_vs_stretch_with_age}(b), where there is an initial sharp increase followed by a plateau.
The convergence of the stretch values between ages 59 and 89 is due to the initial end-to-end length $r_0$ approaching the contour length $L$.
As $r_0$ increases with age, the amount of allowable macroscopic stretch is reduced.

\begin{figure}[htb!]
    \centering
    \includegraphics[width=\textwidth]{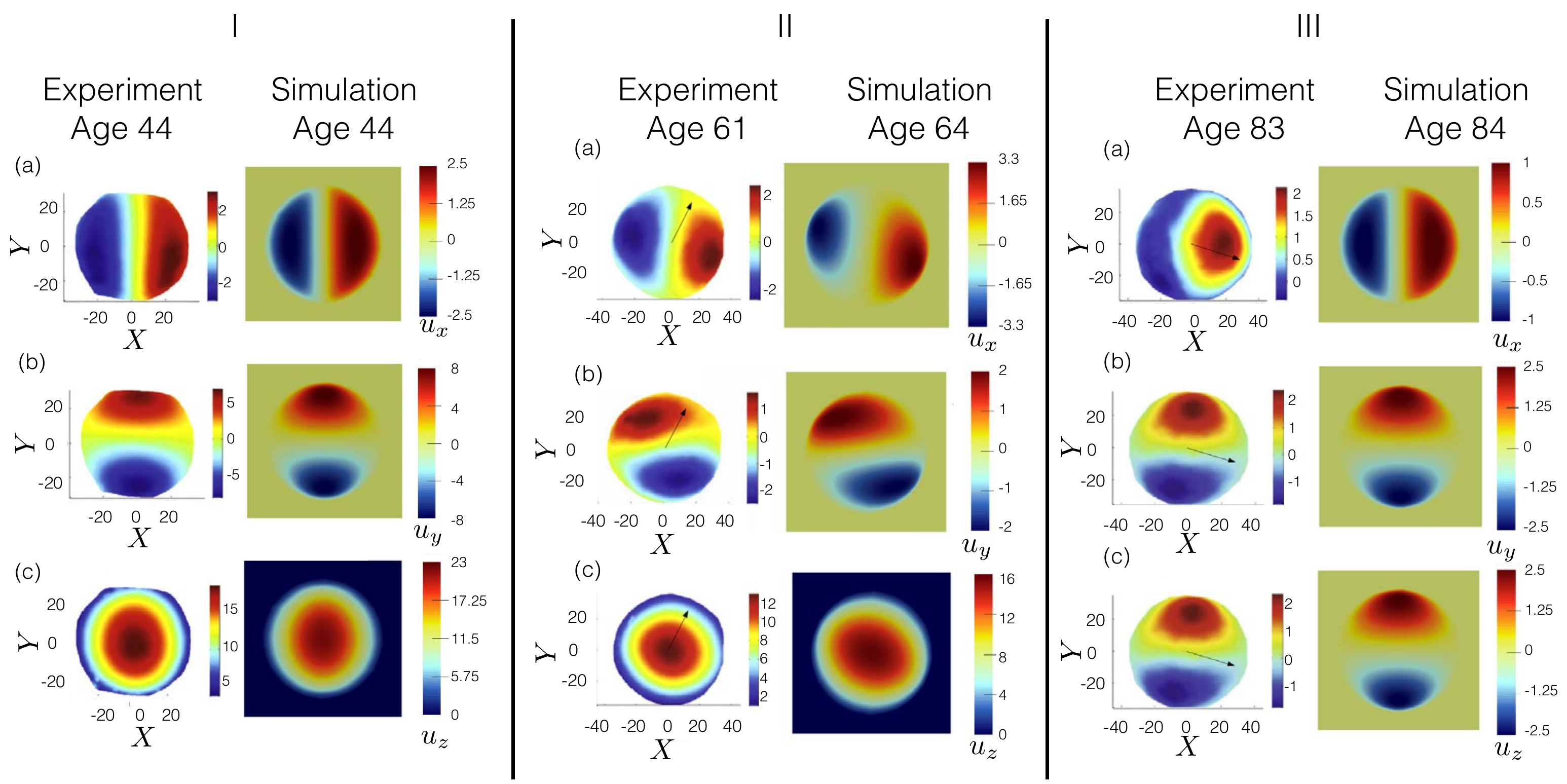}
    \caption{Plots of deformed profile for the bulge test coloured by the various componenets of the displacement [\si{\mm}].
    The simulations are compared to the experimental results of \citet{Tonge2013b} for various different ages (I--III).}
    \label{combined_results}
\end{figure}

The contour plots of the displacement components for the age 44 specimen is displayed in \fig{combined_results}(I), with the fibre direction along the $x$-axis.
A comparison is given with the experimental results obtained by \citet{Tonge2013b}.
As expected, the contour profiles for each displacement component are comparable and the overall behaviour of the simulations matches the experimental results.
The age 44 simulations accurately capture the displacement along the $x$-axis \fig{combined_results}(Ia), but the displacements along the $y$- and $z$-axes \fig{combined_results}(Ib-Ic) are slightly overestimated by approximately \SI{0.5}{\mm} and \SI{2}{\mm}, respectively.
This overestimation arises from a constituent contribution not captured in the model.
It is also possible that the effect of the rig set-up is not appropriately captured in the simulation.
The guard ring may introduce a compressive force through the thickness that would limit the extent of the deformation.

\fig{combined_results}(II-III) show the simulated contour profile of the bulge test for samples according to the proposed ``ageing'' of the various parameters.
These profiles are compared to experimental results at approximately the same age, as extracted from the data obtained by \citet{Tonge2013b}.
It is important to note that in the experimental specimens the Langer lines were not aligned with the body axes.
The Langer lines were oriented at \ang{64} and \ang{-24.6} degrees to the $x$-axis for the age 61 and 83 specimens, respectively, as indicated in the figures.
This alteration in the orientation is reflected in the simulated results.

The profile comparison between the age 64 simulated and age 61 experimental results are very similar to those of age 44.
The alteration in the fibre orientation is captured quite satisfactorily, with the resulting axial assymmetry in the components of displacement along the $x$- and $y$-axes represented accurately.
The $z$ component of displacement in the out-of-plane direction is acceptably captured, although there seems to be an over estimation in the level of anisotropy in the simulated results.
In terms of the magnitude of the displacements, the $y$-axis components of the age 61 experiment and age 64 simulation are similar, but the simulations overestimate the other two components.
This is not too surprising considering the unsatisfactory fit with the pressure-stretch data at age 61 as shown in \fig{p_vs_stretch_with_age}.

The age 84 simulations were orientated at \ang{-24.6} to the positive $x$-axis in order to replicate the age 83 experimental contour plots.
There is sufficient agreement in the general profile of the simulated results when compared to the experimental data.
The only discrepancy lies in the component along the $x$-axis of the experimental data as shown in \fig{combined_results}(IIIa), where the lack of symmetry suggests a possible defect in the skin specimen, such as a non-uniform skin thickness or an irregularity on a constituent level, such as an inconsistent collagen distribution or dispersion.
Despite this, the displacements along the $x$- and $y$-axes of the age 84 simulated results are comparable to the experimental results, but there is again a minor overestimation in the $z$-axis component.

\section{Discussion}\label{sec_discussion}

The impact of ageing on the mechanical response of human skin was investigated using a microstructurally-motivated constitutive continuum description featuring distinct phases representing the ground substance, elastin and collagen.
The ground substance and elastin contributions were modelled by an isotropic, coupled, and compressible neo-Hookean free energy.
For collagen, the primary load-bearing constituent of skin in tension, a wormlike eight-chain model motivated by the fibrous and intertwined nature of the collagen network was selected. This approach captures well the inherent nonlinearity and anisotropy of the skin under finite deformations.

A subset of the constitutive parameters whose age-dependent variation has the most physically justified influence on the response was identified.
The parameters controlled age-related structural changes such as loss of integrity within the elastin network, dermal thinning and unravelling of the collagen network. The continuum model was then used to predict the response of bulge tests which, compared to uniaxial extension tests, provide a richer set of deformations and stress states to evaluate the model performance. The proposed constitutive model was found to adequately capture the mechanical response.

By experimentally-informed modification to the constitutive parameters the ageing process was successfully replicated. The reduction in stiffness at low stretches (low modulus portion) through degradation of the elastin network was well captured through modification of the elastin free energy.
Results show that degradation of the elastin meshwork and variations in anisotropy of the collagen network are plausible mechanisms to explain ageing in terms of macroscopic tissue stiffening. Whereas alterations in elastin affect the low-strain region of the skin stress-strain curve, those related to collagen have an impact on its (large strain) linear region.

As with any modelling study, and despite the ability of the proposed constitutive model to capture ageing effects, some limitations can be identified and form the basis for future improvements.
In order to incorporate an age-based modification to the model microstructural constitutive parameters, experimental bulge test data was used at ages 44, 61 and 83. At only three sample points, this does not represent an adequate range of data with which to make a conclusive parameter fit. Therefore, for a more comprehensive ageing model to be developed, it would be necessary to include data from a statistically significant number of specimens over a large range of age groups. Additionally, experimental samples would need to be controlled for several factors so as to limit variability, or at least, to characterise it so it could be accounted for in the modelling. Of significant importance would be control over anatomical site. Not only should skin from the same site of the patients be tested, but also controlled for segregating the respective influence of intrinsic ageing and photoageing.

The model does not \emph{explicitly} account for crosslinking within the collagen network, and therefore, cannot provide a mechanistic description of this effect.
 Crosslinking prevents slippage between fibres and accordingly contributes toward the stiffness elicited by the collagen network under stretch \citep{Tang2010}.
With age, crosslinking has been observed to increase which has often been attributed as a possible source of the increase in macroscopic stiffness \citep{Quatresooz2006}.
Here, more complex chemomechanobiological processes are at play and increase in crosslinking might rather be a by-product of UV radiation exposure and glycation associated with diabetes \citep{Sherratt2013} rather than the consequence of intrinsic ageing per se.
 Modifications of crosslinking properties are implicitly captured in the eight-chain model.
 For a more explicit description of fibre-to-fibre and matrix-to-fibre mechanical/physical interactions, other approaches are possible.
 One could use dedicated tensor invariants \citep{Limbert2011, Peng2005} that segregate deformation modes associated with such interactions or one could use multiscale mechanistic micromechanical constitutive models that explicitly describe these interactions \citep{Maceri2010, Marino2013, Marino2015}.

The mechanical response of the collagen meshwork was captured using an eight-chain entropic model based on a wormlike chain energy.
This type of model is rooted in statistical mechanics where the assumption is that long chain macromolecular networks can take a very large number of possible conformations.
This assumption might be questionable as it is unclear whether the highly-bonded and hierarchical structure of collagen assemblies allows degrees of freedom for all states to be sampled in an entropy-dominated regime.
Moreover, the wormlike chain is assumed to represent an individual fibre or a bundle of collagen fibres.
At that length scale, it is expected that enthalpic effects would be dominant over entropic effects.
However, energetic elasticity models can also model the typical strain hardening behaviour of biological soft tissues by uncoiling of crimps \citep{Garikipati2008}.
Other types of microstructurally-motivated strain energy function for the individual chains of the unit cell could also be used \citep{Holzapfel2000, Schroder2003}.

The skin was modelled here as single homogeneous layer as this was supported by the reasonable assumption that the dermal layer is the main mechanical contributor of the skin under states of tension.
In order to develop a more realistic model it would be relevant to incorporate a multi-layer model with realistic structural geometries \citep{Limbert2017, LeyvaMendivil2017}.

 It is worth highlighting that, with age, although the skin undergoes drastic alterations of its own biophysical properties, particularly mechanical properties, distinct ageing processes occur in other body tissues, structures and organs.
 They have the potential to alter the mechanical environment experienced by the skin.
 For example, thinning/atrophy of adipose and muscular tissues \citep{Quatresooz2006}, or even bone resorption, are factors that can modify the complex mechanobiology and associated residual strains of the skin as an organ covering the entire body.
 This is consistent with observations that the magnitude and directions of Langer lines \citep{Langer1978} vary with age.
 These aspects were not directly accounted for in the present constitutive model and could explain the inability to obtain an excellent fit between experimental and model data for the inflation tests eliciting the skin response along the main fibre direction (\fig{t_vs_age}).
 Also, with age, the skin tends to lose its in-plane isotropy \citep{Ruvolo2007} because of the strong mechanical effects introduced by dermal collagen realignment arising in combination with collagen cross-linking and density alteration.

In intrinsic skin ageing, the dermal-epidermal junction (DEJ) which is a 0.5-\SI{1}{\micro\m} thick three-dimensional interlocking wavy interface basement membrane \citep{Chan1997} connecting the dermis to the epidermis, tends to flatten out \citep{Lavker1979, Lagarrigue2012, Sauermann2002, Querleux2009, Kligman1985}.
This reduction in the amplitude of the papillae which are the protrusions of the dermis into the epidermis (see \fig{fig_skin_structure}), is also accompanied by a decrease in their density (unit per surface area) \citep{Lagarrigue2012,Sauermann2002}.
This alters the mechanical interactions between the epidermis and dermis resulting in decreased resistance to shear deformation at the DEJ.
In turn, this is likely to modulate the macroscopic mechanical response of the skin as evidenced by the increasing frailty of skin with age \citep{Morey2007}.
Besides these important structural effects, the flattening of the DEJ has important metabolic consequences as it reduces the surface area for nutritional exchange and metabolic by-products evacuation between the dermis and epidermis. As a result, epidermal cell turnover is slowed down and free radicals accumulate.

At this stage, the model does not account for ageing-triggered enzymatic degradation of elastic fibres, abnormal collagen deposition and remodelling.
At the sub-cellular and cellular levels, there exists a variety of enzymes and cytokines that are active in the ageing of the skin’s primary constituents. Intrinsic ageing is accompanied by a reduction of the fibroblast population \citep{Fenske1986} which induces a decrease in collagen production and is associated with an increased build-up of matrix metalloproteinases (MMP) \citep{Varani2001}.
MMP are enzymes that can cleave elastic fibre molecules \citep{Varani2001, Ashworth1999} and are believed to play an active role in degrading important structural proteins in the skin such as collagen, elastin, fibronectin and laminin.
Disruption of the homeostatic state between activation and inhibition of MMPs is a key ingredient in the pathophysiology of intrinsic skin ageing and photoageing by controlling the degradation of collagen and elastin in the skin.
Reactive oxygen species (ROS) are known to play a crucial part in driving the ageing process. Such chemical species could form the building blocks of mixture type models \citep{Humphrey2003, Rouhi2007, Valentin2009, Valentin2012} or models based on open-system thermodynamics \citep{Kuhl2003, Menzel2012}.

The number of processes, their interplay, and the scales at which these occur make deconstructing skin ageing a very challenging task at both the experimental and modelling levels.
Additionally, the effects of intrinsic ageing are generally compounded by the influence of extrinsic ageing (e.g.\ photoageing), which only serves to add complexity to the problem.
In summary, much work remains to be done to develop a continuum description of skin capable of incorporating the full ageing process.
These aspects are currently being implemented in a new chemomechanobiological constitutive model which will be the subject of subsequent publications.

\section{Acknowledgements}

This work was funded through the award of a Royal Society Newton Fund grant (2014-2016) between the Universities of Southampton and Cape Town.
The authors would like to gratefully acknowledge this financial support as well as the logistic and infrastructure support provided by their respective institutions for research visits of DP, AM and GL.

%
%
%
\section{References}
\bibliographystyle{unsrtnat}
\bibliography{refs.bib}

\begin{thebibliography}{119}
\providecommand{\natexlab}[1]{#1}
\providecommand{\url}[1]{\texttt{#1}}
\expandafter\ifx\csname urlstyle\endcsname\relax
  \providecommand{\doi}[1]{doi: #1}\else
  \providecommand{\doi}{doi: \begingroup \urlstyle{rm}\Url}\fi

\bibitem[Jor et~al.(2013)Jor, Parker, Taberner, Nash, and Nielsen]{Jor2013}
J.~W.~Y. Jor, M.~D. Parker, A.~J. Taberner, M.~P. Nash, and P.~M.~F. Nielsen.
\newblock Computational and experimental characterization of skin mechanics:
  identifying current challenges and future directions.
\newblock \emph{Wiley Interdisciplinary Reviews: Systems Biology and Medicine},
  5\penalty0 (5):\penalty0 539--556, 2013.

\bibitem[Li(2015)]{Li2015}
Wenguang Li.
\newblock Modelling methods for in vitro biomechanical properties of the skin:
  {A} review.
\newblock \emph{Biomedical Engineering Letters}, 5\penalty0 (4):\penalty0
  241--250, 2015.

\bibitem[Limbert(2017)]{Limbert2017}
G.~Limbert.
\newblock Mathematical and computational modelling of skin biophysics-a review.
\newblock \emph{Proceedings of the Royal Society A-Mathematical Physical and
  Engineering Sciences}, 473\penalty0 (2203):\penalty0 1--39, 2017.

\bibitem[Silver et~al.(2003)Silver, Siperko, and Seehra]{Silver2003}
F.~H. Silver, L.~M. Siperko, and G.~P. Seehra.
\newblock Mechanobiology of force transduction in dermal tissue.
\newblock \emph{Skin Research and Technology}, 9\penalty0 (1):\penalty0 3--23,
  2003.

\bibitem[web(2010)]{web2010}
\url{www.ons.gov.uk/peoplepopulationandcommunity/birthsdeathsandmarriages/ageing/
  articles/characteristicsofolderpeople/2013-12-06}, 2010.

\bibitem[web(2016)]{web2016}
\url{www.ons.gov.uk/peoplepopulationandcommunity/populationandmigration/populationestimates/
  articles/overviewoftheukpopulation/february2016}, 2016.

\bibitem[Morey(2007)]{Morey2007}
P.~Morey.
\newblock Skin tears: a literature review.
\newblock \emph{Primary Intention}, 15\penalty0 (3):\penalty0 122--129, 2007.

\bibitem[Shimizu(2007)]{Shimizu2007}
H.~Shimizu.
\newblock \emph{Shimizu's Textbook of Dermatology}.
\newblock Hokkaido University Press - Nakayama Shoten Publishers, 2007.

\bibitem[Fung(1981)]{Fung1981}
Y.~C. Fung.
\newblock \emph{Biomechanics: Mechanical Properties of Living Tissues}.
\newblock Springer-Verlag, New York, 1981.

\bibitem[Limbert(2014)]{Limbert2014}
G.~Limbert.
\newblock \emph{Chapter 4: State-of-the-art constitutive models of skin
  biomechanics}, chapter~4, pages 95--131.
\newblock Pan Stanford Publishing Pte. Ltd, Singapore, 2014.

\bibitem[Lanir(1987)]{Lanir1987}
Y.~Lanir.
\newblock \emph{Skin mechanics}.
\newblock McGraw-Hill, New York, 1987.

\bibitem[Ramos-e Silva and da~Silva~Carneiro(2007)]{Ramos2007}
M.~Ramos-e Silva and S.~C. da~Silva~Carneiro.
\newblock Elderly skin and its rejuvenation: products and procedures for the
  aging skin.
\newblock \emph{Journal of Cosmetic Dermatology}, 6\penalty0 (1):\penalty0
  40--50, 2007.

\bibitem[Assaf et~al.(2010)Assaf, Adly, and Hussein]{Assaf2014}
H.~Assaf, M.~A. Adly, and M.~R. Hussein.
\newblock \emph{Aging and Intrinsic Aging: Pathogenesis and Manifestations},
  book section~13, pages 129--138.
\newblock 2010.

\bibitem[Naylor et~al.(2011)Naylor, Watson, and Sherratt]{Naylor2011}
E.~C. Naylor, R.~E.~B. Watson, and M.~J. Sherratt.
\newblock Molecular aspects of skin ageing.
\newblock \emph{Maturitas}, 69\penalty0 (3):\penalty0 249--256, 2011.

\bibitem[Gilchrest and Yaar(1992)]{Gilchrest1992}
B.~A. Gilchrest and M.~Yaar.
\newblock Ageing and photoageing of the skin: observations at the cellular and
  molecular level.
\newblock \emph{British Journal of Dermatology}, 127 Suppl 41:\penalty0 25--30,
  1992.

\bibitem[Goukassian et~al.(2000)Goukassian, Gad, Yaar, Eller, Nehal, and
  Gilchrest]{Goukassian2000}
D.~Goukassian, F.~Gad, M.~Yaar, M.~S. Eller, U.~S. Nehal, and B.~A. Gilchrest.
\newblock Mechanisms and implications of the age-associated decrease in dna
  repair capacity.
\newblock \emph{FASEB J}, 14\penalty0 (10):\penalty0 1325--34, 2000.

\bibitem[Giacomoni and Rein(2004)]{Giacomoni2004}
P.~U. Giacomoni and G.~Rein.
\newblock A mechanistic model for the aging of human skin.
\newblock \emph{Micron}, 35\penalty0 (3):\penalty0 179--84, 2004.

\bibitem[Fisher et~al.(1997)Fisher, Wang, Datta, Varani, Kang, and
  Voorhees]{Fisher1997}
G.~J. Fisher, Z.~Wang, S.~C. Datta, J.~Varani, S.~Kang, and J.~J. Voorhees.
\newblock Pathophysiology of premature skin aging induced by ultraviolet light.
\newblock \emph{New England Journal of Medicine}, 337\penalty0 (20):\penalty0
  1419--1429, 1997.

\bibitem[Berneburg et~al.(2000)Berneburg, Plettenberg, and
  Krutmann]{Berneburg2000}
M.~Berneburg, H.~Plettenberg, and J.~Krutmann.
\newblock Photoaging of human skin.
\newblock \emph{Photodermatol Photoimmunol Photomed}, 16\penalty0 (6):\penalty0
  239--44, 2000.

\bibitem[Kligman(1969)]{Kligman1969}
A.~M. Kligman.
\newblock Early destructive effects of sunlight on human skin.
\newblock \emph{Journal of the American Medical Association}, 210:\penalty0
  2377--2380, 1969.

\bibitem[Fisher et~al.(2002)Fisher, Kang, Varani, Bata-Csorgo, Wan, Datta, and
  Voorhees]{Fisher2002}
G.~J. Fisher, S.~Kang, J.~Varani, Z.~Bata-Csorgo, Y.~Wan, S.~Datta, and J.~J.
  Voorhees.
\newblock Mechanisms of photoaging and chronological skin aging.
\newblock \emph{Archives of Dermatology}, 138\penalty0 (11):\penalty0 1462--70,
  2002.

\bibitem[Diffey(2003)]{Diffey2003}
B.~L. Diffey.
\newblock A quantitative estimate of melanoma mortality from ultraviolet a
  sunbed use in the uk.
\newblock \emph{British Journal of Dermatology}, 149\penalty0 (3):\penalty0
  578--581, 2003.

\bibitem[Sandby-Moller et~al.(2003)Sandby-Moller, Poulsen, and
  Wulf]{SandbyMoller2003}
J.~Sandby-Moller, T.~Poulsen, and H.~C. Wulf.
\newblock Epidermal thickness at different body sites: relationship to age,
  gender, pigmentation, blood content, skin type and smoking habits.
\newblock \emph{Acta Dermato Venereologica}, 83\penalty0 (6):\penalty0 410--3,
  2003.

\bibitem[Vierkötter and Krutmann(2012)]{Vierkotter2012}
A.~Vierkötter and J.~Krutmann.
\newblock Environmental influences on skin aging and ethnic-specific
  manifestations.
\newblock \emph{Dermato-endocrinology}, 4\penalty0 (3):\penalty0 227--231,
  2012.

\bibitem[Benedetto(1998)]{Benedetto1998}
A.~V. Benedetto.
\newblock The environment and skin aging.
\newblock \emph{Clinical Dermatology}, 16\penalty0 (1):\penalty0 129--139,
  1998.

\bibitem[Pawlaczyk et~al.(2013)Pawlaczyk, Lelonkiewicz, and
  Wieczorowski]{Pawlaczyk2013}
M.~Pawlaczyk, M.~Lelonkiewicz, and M.~Wieczorowski.
\newblock Age-dependent biomechanical properties of the skin.
\newblock \emph{Postepy Dermatologii i Alergologii}, 30\penalty0 (5):\penalty0
  302--306, 2013.

\bibitem[Sherratt(2013)]{Sherratt2013}
M.~J. Sherratt.
\newblock Age-related tissue stiffening: Cause and effect.
\newblock \emph{Advances in Wound Care}, 2\penalty0 (1):\penalty0 11--17, 2013.

\bibitem[Silver et~al.(2002)Silver, Seehra, Freeman, and DeVore]{Silver2002}
F.~H. Silver, G.~Seehra, J.~W. Freeman, and D.~DeVore.
\newblock Viscoelastic of young and old human dermis: a proposed moelcular
  mechanism for elastic energy storage in collagen and elastin.
\newblock \emph{Journal of Applied Polymer Science}, 86:\penalty0 1978--1985,
  2002.

\bibitem[Ruvolo~Jr et~al.(2007)Ruvolo~Jr, Stamatas, and Kollias]{Ruvolo2007}
E.~C. Ruvolo~Jr, G.~N. Stamatas, and N.~Kollias.
\newblock Skin viscoelasticity displays site- and age-dependent angular
  anisotropy.
\newblock \emph{Skin Pharmacology and Physiology}, 20\penalty0 (6):\penalty0
  313--321, 2007.

\bibitem[Leyva-Mendivil et~al.(2017{\natexlab{a}})Leyva-Mendivil, Lengiewicz,
  Page, Bressloff, and Limbert]{LeyvaMendivil2017}
M.~F. Leyva-Mendivil, J.~Lengiewicz, A.~Page, N.~W. Bressloff, and G.~Limbert.
\newblock Skin microstructure is a key contributor to its friction behaviour.
\newblock \emph{Tribology Letters}, 65\penalty0 (1):\penalty0 12,
  2017{\natexlab{a}}.

\bibitem[Leyva-Mendivil et~al.(2017{\natexlab{b}})Leyva-Mendivil, Lengiewicz,
  Page, Bressloff, and Limbert]{LeyvaMendivil2017a}
M.~F. Leyva-Mendivil, J.~Lengiewicz, A.~Page, N.~W. Bressloff, and G.~Limbert.
\newblock Implications of multi-asperity contact for shear stress distribution
  in the viable epidermis - an image-based finite element study.
\newblock \emph{Biotribology}, page in press, 2017{\natexlab{b}}.

\bibitem[Leyva-Mendivil et~al.(2015)Leyva-Mendivil, Page, Bressloff, and
  Limbert]{LeyvaMendivil2015}
M.~F. Leyva-Mendivil, A.~Page, N.~W. Bressloff, and G~Limbert.
\newblock A mechanistic insight into the mechanical role of the stratum corneum
  during stretching and compression of the skin.
\newblock \emph{Journal of the Mechanical Behavior of Biomedical Materials},
  49:\penalty0 197--219, 2015.

\bibitem[Hahnel et~al.(2017)Hahnel, Lichterfeld, Blume-Peytavi, and
  Kottner]{Hahnel2017}
E.~Hahnel, A.~Lichterfeld, U.~Blume-Peytavi, and J.~Kottner.
\newblock The epidemiology of skin conditions in the aged: A systematic review.
\newblock \emph{Journal of Tissue Viability}, 26\penalty0 (1):\penalty0 20--28,
  2017.

\bibitem[Goriely and Ben~Amar(2007)]{Goriely2007}
A.~Goriely and M.~Ben~Amar.
\newblock On the definition and modeling of incremental, cumulative, and
  continuous growth laws in morphoelasticity.
\newblock \emph{Biomechanics and Modeling in Mechanobiology}, 6\penalty0
  (5):\penalty0 289--296, 2007.

\bibitem[Ambrosi et~al.(2011)Ambrosi, Ateshian, Arruda, Cowin, Dumais, Goriely,
  Holzapfel, Humphrey, Kemkemer, Kuhl, Olberding, Taber, and
  Garikipati]{Ambrosi2011}
D.~Ambrosi, G.~A. Ateshian, E.~M. Arruda, S.~C. Cowin, J.~Dumais, A.~Goriely,
  G.~A. Holzapfel, J.~D. Humphrey, R.~Kemkemer, E.~Kuhl, J.~E. Olberding, L.~A.
  Taber, and K.~Garikipati.
\newblock Perspectives on biological growth and remodeling.
\newblock \emph{Journal of the Mechanics and Physics of Solids}, 59\penalty0
  (4):\penalty0 863--883, 2011.

\bibitem[Xu and Lu(2011)]{Xu2011}
F.~Xu and T.~Lu.
\newblock \emph{Introduction to Skin Biothermomechanics and Thermal Pain}.
\newblock Springer, Heidelberg Dordrecht London New York, 2011.

\bibitem[Flynn(2014)]{Flynn2014}
C.~Flynn.
\newblock \emph{Fiber-matrix models of the dermis}, book section~5, pages
  133--159.
\newblock Pan Stanford Publishing Pty Ltd, Singapore, 2014.

\bibitem[Kuhl et~al.(2005)Kuhl, Garikipati, Arruda, and Grosh]{Kuhl2005}
E.~Kuhl, K.~Garikipati, E.~Arruda, and K.~Grosh.
\newblock Remodeling of biological tissue: Mechanically induced reorientation
  of a transversely isotropic chain network.
\newblock \emph{Journal of the Mechanics and Physics of Solids}, 53:\penalty0
  1552--1573, 2005.

\bibitem[Kuhl and Holzapfel(2007)]{Kuhl2007}
E.~Kuhl and G.~A. Holzapfel.
\newblock A continuum model for remodeling in living structures.
\newblock \emph{Journal of Materials Science}, 42\penalty0 (21):\penalty0
  8811--8823, 2007.
\newblock ISSN 0022-2461.

\bibitem[Garikipati et~al.(2004)Garikipati, Arruda, Grosh, Narayanan, and
  Calve]{Garikipati2004}
K.~Garikipati, E.~M. Arruda, K.~Grosh, H.~Narayanan, and S.~Calve.
\newblock A continuum treatment of growth in biological tissue: the coupling of
  mass transport and mechanics.
\newblock \emph{Journal of the Mechanics and Physics of Solids}, 52\penalty0
  (7):\penalty0 1595--1625, 2004.

\bibitem[Mazza et~al.(2005)Mazza, Papes, Rubin, Bodner, and Binur]{Mazza2005}
E.~Mazza, O.~Papes, M.~B. Rubin, S.~R. Bodner, and N.~S. Binur.
\newblock Nonlinear elastic-viscoplastic constitutive equations for aging
  facial tissues.
\newblock \emph{Biomechanics and Modeling in Mechanobiology}, 4\penalty0
  (2-3):\penalty0 178--189, 2005.

\bibitem[Mazza et~al.(2007)Mazza, Papes, Rubin, Bodner, and Binur]{Mazza2007}
E.~Mazza, O.~Papes, M.~B. Rubin, S.~R. Bodner, and N.~S. Binur.
\newblock Simulation of the aging face.
\newblock \emph{Journal of Biomechanical Engineering-Transactions of the Asme},
  129\penalty0 (4):\penalty0 619--623, 2007.

\bibitem[Rubin and Bodner(2002)]{Rubin2002}
M.~B. Rubin and S.~R. Bodner.
\newblock A three-dimensional nonlinear model for dissipative response of soft
  tissue.
\newblock \emph{International Journal of Solids and Structures}, 39\penalty0
  (19):\penalty0 5081--5099, 2002.

\bibitem[Maceri et~al.(2013)Maceri, Marino, and Vairo]{Maceri2013}
F.~Maceri, M.~Marino, and G.~Vairo.
\newblock Age-dependent arterial mechanics via a multiscale elastic approach.
\newblock \emph{International Journal for Computational Methods in Engineering
  Science and Mechanics}, 14\penalty0 (2):\penalty0 141--151, 2013.

\bibitem[Tonge et~al.(2013{\natexlab{a}})Tonge, Atlan, Voo, and
  Nguyen]{Tonge2013}
T.~K. Tonge, L.~S. Atlan, L.~M. Voo, and T.~D. Nguyen.
\newblock Full-field bulge test for planar anisotropic tissues: {P}art {I} –
  experimental methods applied to human skin tissue.
\newblock \emph{Acta Biomaterialia}, 9\penalty0 (4):\penalty0 5913--5925,
  2013{\natexlab{a}}.

\bibitem[Tonge et~al.(2013{\natexlab{b}})Tonge, Voo, and Nguyen]{Tonge2013b}
T.~K. Tonge, L.~M. Voo, and T.~D. Nguyen.
\newblock Full-field bulge test for planar anisotropic tissues: {P}art {II} –
  a thin shell method for determining material parameters and comparison of two
  distributed fiber modeling approaches.
\newblock \emph{Acta Biomaterialia}, 9\penalty0 (4):\penalty0 5926--5942,
  2013{\natexlab{b}}.

\bibitem[Silver et~al.(2001)Silver, Freeman, and DeVore]{Silver2001}
F.~H. Silver, J.~W. Freeman, and D.~DeVore.
\newblock Viscoelastic properties of human skin and processed dermis.
\newblock \emph{Skin Research and Technology}, 7\penalty0 (1):\penalty0 18--23,
  2001.

\bibitem[L\'{e}v\^{e}que et~al.(1980)L\'{e}v\^{e}que, de~Rigal, Agache, and
  Monneur]{Leveque1980}
J.~L. L\'{e}v\^{e}que, J.~de~Rigal, P.~G. Agache, and C.~Monneur.
\newblock Influence of ageing on the in vivo extensibility of human skin at a
  low stress.
\newblock \emph{Archives of Dermatological Research}, 269\penalty0
  (2):\penalty0 127--135, 1980.

\bibitem[Reihsner et~al.(1995)Reihsner, Balogh, and Menzel]{Reihsner1995}
R.~Reihsner, B.~Balogh, and E.~J. Menzel.
\newblock Two-dimensional elastic properties of human skin in terms of an
  incremental model at the in vivo configuration.
\newblock \emph{Medical Engineering and Physics}, 17\penalty0 (4):\penalty0
  304--313, 1995.

\bibitem[Oxlund and Andreassen(1980)]{Oxlund1980}
H.~Oxlund and T.~T. Andreassen.
\newblock The roles of hyaluronic acid, collagen and elastin in the mechanical
  properties of connective tissues.
\newblock \emph{Journal of Anatomy}, 131\penalty0 (4):\penalty0 611--620, 1980.

\bibitem[Oxlund et~al.(1988)Oxlund, Manschot, and Viidik]{Oxlund1988}
H.~Oxlund, J.~Manschot, and A.~Viidik.
\newblock The role of elastin in the mechanical properties of skin.
\newblock \emph{Journal of Biomechanics}, 21\penalty0 (3):\penalty0 213--218,
  1988.

\bibitem[Pailler-Mattei et~al.(2008)Pailler-Mattei, Bec, and
  Zahouani]{PaillerMattei2008}
C.~Pailler-Mattei, S.~Bec, and H.~Zahouani.
\newblock In vivo measurements of the elastic mechanical properties of human
  skin by indentation tests.
\newblock \emph{Medical Engineering \& Physics}, 30\penalty0 (5):\penalty0
  599--606, 2008.

\bibitem[Ribeiro et~al.(2013)Ribeiro, {dos Anjos}, Mello, and
  de~Campos~Vidal]{Ribeiro2013}
J.~F. Ribeiro, E.~H.~M. {dos Anjos}, Maria Luiza~S. Mello, and
  B.~de~Campos~Vidal.
\newblock Skin collagen fiber molecular order: a pattern of distributional
  fiber orientation as assessed by optical anisotropy and image analysis.
\newblock \emph{PLOS ONE}, 8\penalty0 (1):\penalty0 e54724, 2013.

\bibitem[Gosline et~al.(2002)Gosline, Lillie, Carrington, Guerette, Ortlepp,
  and Savage]{Gosline2002}
J.~Gosline, M.~Lillie, E.~Carrington, P.~Guerette, C.~Ortlepp, and K.~Savage.
\newblock Elastic proteins: biological roles and mechanical properties.
\newblock \emph{Philosophical Transactions of the Royal Society of London.
  Series B: Biological Sciences}, 357\penalty0 (1418):\penalty0 121--132, 2002.

\bibitem[Langer(1978)]{Langer1978}
K.~Langer.
\newblock On the anatomy and physiology of the skin: {II}. {S}kin tension.
\newblock \emph{British Journal of Plastic Surgery}, 31\penalty0 (2):\penalty0
  93--106, 1978.

\bibitem[Lapeer et~al.(2010)Lapeer, Gasson, and Karri]{Lapeer2010}
R.~J. Lapeer, P.~D. Gasson, and V.~Karri.
\newblock Simulating plastic surgery: From human skin tensile tests, through
  hyperelastic finite element models to real-time haptics.
\newblock \emph{Progress in Biophysics and Molecular Biology}, 103\penalty0
  (2-3):\penalty0 208--216, 2010.

\bibitem[Silver et~al.(1992)Silver, Kato, Ohno, and Wasserman]{Silver1992}
F.~H. Silver, Y.~P. Kato, M.~Ohno, and A.~J. Wasserman.
\newblock Analysis of mammalian connective tissue: relationship between
  hierarchical structures and mechanical properties.
\newblock \emph{Journal of long-term effects of medical implants}, 2\penalty0
  (2-3):\penalty0 165--198, 1992.

\bibitem[Tregear(1969)]{Tregear1969}
R.~T. Tregear.
\newblock The mechanical properties of skin.
\newblock \emph{Journal of the Society of Cosmetic Chemists}, 20:\penalty0
  467--477, 1969.

\bibitem[Oomens et~al.(1987)Oomens, van Campen, and Grootenboer]{Oomens1987}
C.~W. Oomens, D.~H. van Campen, and H.~J. Grootenboer.
\newblock In vitro compression of a soft tissue layer on a rigid foundation.
\newblock \emph{Journal of Biomechanics}, 20\penalty0 (10):\penalty0 923--935,
  1987.

\bibitem[Escoffier et~al.(1989)Escoffier, {de Rigal}, Rochefort, Vasselet,
  Lévêque, and Agache]{Escoffier1989}
C.~Escoffier, J.~{de Rigal}, A.~Rochefort, R.~Vasselet, J.~L. Lévêque, and
  P.~G. Agache.
\newblock Age-related mechanical properties of human skin: an in vivo study.
\newblock \emph{The Journal of Investigative Dermatology}, 93\penalty0
  (3):\penalty0 353--357, 1989.

\bibitem[Diridollou et~al.(2001)Diridollou, Vabre, Berson, Vaillant, Black,
  Lagarde, Gregoire, Gall, and Patat]{Diridollou2001}
S.~Diridollou, V.~Vabre, M.~Berson, L.~Vaillant, D.~Black, J.~M. Lagarde, J.~M.
  Gregoire, Y.~Gall, and F.~Q. Patat.
\newblock Skin ageing: changes of physical properties of human skin in vivo.
\newblock \emph{International Journal of Cosmetic Science}, 23\penalty0
  (6):\penalty0 353--62, 2001.

\bibitem[Oriba et~al.(1996)Oriba, Bucks, and Maibach]{Oriba1996}
H.~A. Oriba, D.~A. Bucks, and H.~I. Maibach.
\newblock Percutaneous absorption of hydrocortisone and testosterone on the
  vulva and foreaarm: effect of the menopause and site.
\newblock \emph{British Journal of Dermatology}, 134:\penalty0 229--233, 1996.

\bibitem[Duncan and Lefell(1997)]{Duncan1997}
K.~O. Duncan and D.~J. Lefell.
\newblock Preoperative assessment of eth elderly patient.
\newblock \emph{Dermatologic Clinics}, 15:\penalty0 583--593, 1997.

\bibitem[Brincat et~al.(1987)Brincat, Kabalan, Stud, Moniz, de~Trafford, and
  Montgomery]{Brincat1987}
M.~P. Brincat, S.~Kabalan, J.~W. Stud, C.~F. Moniz, J.~de~Trafford, and
  J.~Montgomery.
\newblock A study of the decrease of skin collagen content, skin thickness, and
  bone mass in the postmeopausal women.
\newblock \emph{Obstretic Gynecology}, 70\penalty0 (6):\penalty0 840--845,
  1987.

\bibitem[Alexander and Cook(2006)]{Alexander2006}
H.~Alexander and T.~Cook.
\newblock Variations with age in the mechanical properties of human skin in
  vivo.
\newblock \emph{Journal of Tissue Viability}, 16\penalty0 (3):\penalty0 6--11,
  2006.

\bibitem[Henry et~al.(1997)Henry, Pi\'{e}rard-Franchimont, Cauwenbergh, and
  Pi\'{e}rard]{Henry1997}
F.~Henry, C.~Pi\'{e}rard-Franchimont, G.~Cauwenbergh, and G.~E. Pi\'{e}rard.
\newblock Age-related changes in facial skin contours and rheology.
\newblock \emph{Journal of the American Geriatrics Society}, 45\penalty0
  (2):\penalty0 220--222, 1997.

\bibitem[Daly and Odland(1979)]{Daly1979}
Colin~H. Daly and George~F. Odland.
\newblock Age-related changes in the mechanical properties of human skin.
\newblock \emph{Journal of Investigative Dermatology}, 73\penalty0
  (1):\penalty0 84--87, 1979.

\bibitem[Chung et~al.(2000)Chung, Kang, Varani, Lin, Fisher, and
  Voorhees]{Chung2000}
J.~H. Chung, S.~Kang, J.~Varani, J.~Lin, G.~J. Fisher, and J.~J. Voorhees.
\newblock Decreased extracellular-signal-regulated kinase and increased
  stress-activated map kinase activities in aged human skin in vivo.
\newblock \emph{Journal of Investigative Dermatology}, 115\penalty0
  (2):\penalty0 177--182, 2000.

\bibitem[Jenkins(2002)]{Jenkins2002}
G.~Jenkins.
\newblock Molecular mechanisms of skin ageing.
\newblock \emph{Mechanisms of Ageing and Development}, 123\penalty0
  (7):\penalty0 801--10, 2002.

\bibitem[Agache et~al.(1980)Agache, Monneur, L\'{e}v\^{e}que, and {de
  Rigal}]{Agache1980}
P.~G. Agache, C.~Monneur, J.~L. L\'{e}v\^{e}que, and J.~{de Rigal}.
\newblock Mechanical properties and {Y}oung's modulus of human skin in vivo.
\newblock \emph{Archives of Dermatological Research}, 269:\penalty0 221--232,
  1980.

\bibitem[Vexler et~al.(1999)Vexler, Polyansky, and Gorodetsky]{Vexler1999}
A.~Vexler, I.~Polyansky, and R.~Gorodetsky.
\newblock Evaluation of skin viscoelasticity and anisotropy by measurement of
  speed of shear wave propagation with viscoelasticity skin analyzer.
\newblock \emph{Journal of Investigative Dermatology}, 113\penalty0
  (5):\penalty0 732--739, 1999.

\bibitem[Holzapfel(2000)]{Holzapfel2000Book}
G.~A. Holzapfel.
\newblock \emph{Nonlinear Solid Mechanics. A Continuum Approach for
  Engineering}.
\newblock John Wiley \& Sons, Chichester, UK, 2000.

\bibitem[Marsden and Hughes(1994)]{Marsden1984}
J.~E. Marsden and T.~J.~R. Hughes.
\newblock \emph{Mathematical Foundations of Elasticity}.
\newblock Dover, New-York, 1994.

\bibitem[Boehler()]{Boehler1978}
J.~P. Boehler.
\newblock Lois de comportement anisotrope des milieux continus.
\newblock \emph{Journal de M\'{e}canique}, \penalty0 (17):\penalty0 153--190.

\bibitem[Spencer(1992)]{Spencer1992}
A.~J.~M. Spencer.
\newblock \emph{Continuum theory of the mechanics of fibre-reinforced
  composites}.
\newblock Springer-Verlag, New York, 1992.

\bibitem[Limbert and Taylor(2002)]{Limbert2002}
G.~Limbert and M.~Taylor.
\newblock On the constitutive modeling of biological soft connective tissues. a
  general theoretical framework and tensors of elasticity for strongly
  anisotropic fiber-reinforced composites at finite strain.
\newblock \emph{International Journal of Solids and Structures}, 39\penalty0
  (8):\penalty0 2343--2358, 2002.

\bibitem[Schr\"{o}der and Neff(2003)]{Schroder2003}
J.~Schr\"{o}der and P.~Neff.
\newblock Invariant formulation of hyperelastic transverse isotropy based on
  polyconvex free energy functions.
\newblock \emph{International Journal of Solids and Structures}, 40\penalty0
  (2):\penalty0 401--445, 2003.

\bibitem[Destrade et~al.(2013)Destrade, MacDonald, Murphy, and
  Saccomandi]{Destrade2013}
M.~Destrade, B.~MacDonald, J.~G. Murphy, and G.~Saccomandi.
\newblock At least three invariants are necessary to model the mechanical
  response of incompressible, transversely isotropic materials.
\newblock \emph{Computational Mechanics}, 52\penalty0 (4):\penalty0 959--969,
  2013.

\bibitem[Arruda and Boyce(1993)]{Arruda1993}
E.~M. Arruda and M.~C. Boyce.
\newblock A three-dimensional constitutive model for the large stretch behavior
  of rubber elastic-materials.
\newblock \emph{Journal of the Mechanics and Physics of Solids}, 41\penalty0
  (2):\penalty0 389--412, 1993.

\bibitem[Flory(1969)]{Flory1969}
P.~J. Flory.
\newblock \emph{Statistical mechanics of chain molecules}.
\newblock John Wiley \& Sons, Chichester-New York, 1969.

\bibitem[Bischoff et~al.(2002)Bischoff, Arruda, and Grosh]{Bischoff2002}
J.~E. Bischoff, E.~A. Arruda, and K.~Grosh.
\newblock A microstructurally based orthotropic hyperelastic constitutive law.
\newblock \emph{Journal of Applied Mechanics-Transactions of the ASME},
  69\penalty0 (5):\penalty0 570--579, 2002.

\bibitem[Kratky and Porod(1949)]{Kratky1949}
O.~Kratky and G.~Porod.
\newblock R\"{o}ntgenuntersuchungen gel\"{o}ster fadenmolek\"{u}le.
\newblock \emph{Recueil des Travaux Chimiques des Pays-Bas et de la Belgique},
  68:\penalty0 1106--1122, 1949.

\bibitem[Marko and Siggia(1995)]{Marko1995}
J.~F. Marko and E.~D. Siggia.
\newblock Stretching dna.
\newblock \emph{Macromolecules}, 28\penalty0 (26):\penalty0 8759--8770, 1995.

\bibitem[{Buganza Tepole} et~al.(2012){Buganza Tepole}, Gosain, and
  Kuhl]{Buganza2012}
A.~{Buganza Tepole}, A.~K. Gosain, and E.~Kuhl.
\newblock Stretching skin: The physiological limit and beyond.
\newblock \emph{International Journal of Non-Linear Mechanics}, 47\penalty0
  (8):\penalty0 938--949, 2012.

\bibitem[Flynn et~al.(2013)Flynn, Taberner, Nielsen, and Fels]{Flynn2013}
C.~Flynn, A.~J. Taberner, P.~M.~F. Nielsen, and S.~Fels.
\newblock Simulating the three-dimensional deformation of in vivo facial skin.
\newblock \emph{Journal of the Mechanical Behavior of Biomedical Materials},
  28\penalty0 (0):\penalty0 484--494, 2013.

\bibitem[Flynn and McCormack(2008)]{Flynn2008}
C.~Flynn and B.~A.~O. McCormack.
\newblock A simplified model of scar contraction.
\newblock \emph{Journal of Biomechanics}, 41\penalty0 (7):\penalty0 1582--1589,
  2008.

\bibitem[Flynn and McCormack(2009)]{Flynn2009}
C.~O. Flynn and B.~A.~O. McCormack.
\newblock A three-layer model of skin and its application in simulating
  wrinkling.
\newblock \emph{Computer Methods in Biomechanics and Biomedical Engineering},
  12\penalty0 (2):\penalty0 125--134, 2009.

\bibitem[Bischoff et~al.(2004)Bischoff, Arruda, and Grosh]{Bischoff2004}
J.~E. Bischoff, E.~M. Arruda, and K.~Grosh.
\newblock A rheological network model for the continuum anisotropic and
  viscoelastic behavior of soft tissue.
\newblock \emph{Biomechanics and Modeling in Mechanobiology}, 3\penalty0
  (1):\penalty0 56--65, 2004.

\bibitem[S\'{a}ez et~al.(2013)S\'{a}ez, Peña, Martínez, and Kuhl]{Saez2013}
P.~S\'{a}ez, E.~Peña, M.~A. Martínez, and E.~Kuhl.
\newblock Computational modeling of hypertensive growth in the human carotid
  artery.
\newblock \emph{Computational Mechanics}, 53\penalty0 (6):\penalty0 1183--1196,
  2013.

\bibitem[Kuhn(1936)]{Kuhn1936}
W.~Kuhn.
\newblock Beziehungen zwischen {M}olekühlgrösse, statistischer
  {M}olek\"{u}lgestalt und elastischen eigenschaften hochpolymerer {S}toffe.
\newblock \emph{Kolloid-Zeitschrift,}, 76:\penalty0 258--271, 1936.

\bibitem[Kuhl et~al.(2003)Kuhl, Menzel, and Steinmann]{Kuhl2003}
E.~Kuhl, A.~Menzel, and P.~Steinmann.
\newblock Computational modeling of growth - {A} critical review, a
  classification of concepts and two new consistent approaches.
\newblock \emph{Computational Mechanics}, 32\penalty0 (1-2):\penalty0 71--88,
  2003.

\bibitem[Kielty et~al.(2002)Kielty, Sherratt, and Shuttleworth]{Kielty2002}
C.~M. Kielty, M.~J. Sherratt, and C.~A. Shuttleworth.
\newblock Elastic fibres.
\newblock \emph{Journal of Cell Science}, 115\penalty0 (14):\penalty0
  2817--2828, 2002.

\bibitem[Korelc(2002)]{Korelc2002}
J.~Korelc.
\newblock Multi-language and multi-environment generation of nonlinear finite
  element codes.
\newblock \emph{Engineering with Computers}, 18\penalty0 (4):\penalty0
  312--327, 2002.

\bibitem[Korelc and Wriggers(2016)]{Korelc2016}
J.~Korelc and P.~Wriggers.
\newblock \emph{Automation of Finite Element Methods}.
\newblock Springer, first edition, 2016.

\bibitem[Jor et~al.(2011)Jor, Nash, Nielsen, and Hunter]{Jor2011}
J.~W.~Y. Jor, M.~P. Nash, P.~M.~F. Nielsen, and P.~J. Hunter.
\newblock Estimating material parameters of a structurally based constitutive
  relation for skin mechanics.
\newblock \emph{Biomechanics and Modeling in Mechanobiology}, 10\penalty0
  (5):\penalty0 767--778, 2011.

\bibitem[Kvistedal and Nielsen(2009)]{Kvistedal2009}
Y.~A. Kvistedal and P.~M.~F. Nielsen.
\newblock Estimating material parameters of human skin in vivo.
\newblock \emph{Biomechanics and Modeling in Mechanobiology}, 8\penalty0
  (1):\penalty0 1--8, 2009.

\bibitem[Tang et~al.(2010)Tang, Ballarini, Buehler, and Eppell]{Tang2010}
Y.~Tang, R.~Ballarini, M.~J. Buehler, and S.~J. Eppell.
\newblock Deformation micromechanisms of collagen fibrils under uniaxial
  tension.
\newblock \emph{Journal of the Royal Society Interface}, 7\penalty0
  (46):\penalty0 839--850, 2010.

\bibitem[Quatresooz et~al.(2006)Quatresooz, Thirion, Pi\'{e}rard-Franchimont,
  and Pi\'{e}rard]{Quatresooz2006}
P.~Quatresooz, L.~Thirion, C.~Pi\'{e}rard-Franchimont, and G.~E. Pi\'{e}rard.
\newblock The riddle of genuine skin microrelief and wrinkles.
\newblock \emph{International Journal of Cosmetic Science}, 28\penalty0
  (6):\penalty0 389--395, 2006.

\bibitem[Limbert(2011)]{Limbert2011}
G.~Limbert.
\newblock A mesostructurally-based anisotropic continuum model for biological
  soft tissues--decoupled invariant formulation.
\newblock \emph{Journal of the Mechanical Behavior of Biomedical Materials},
  4\penalty0 (8):\penalty0 1637--1657, 2011.

\bibitem[Peng et~al.(2005)Peng, Guo, and Moran]{Peng2005}
X.~Q. Peng, Z.~Y. Guo, and B.~Moran.
\newblock An anisotropic hyperelasticconstitutive model with fiber-matrix shear
  interaction for the human annulus fibrosus.
\newblock \emph{Journal of Applied Mechanics}, 73\penalty0 (5):\penalty0
  815--824, 2005.

\bibitem[Maceri et~al.(2010)Maceri, Marino, and Vairo]{Maceri2010}
F.~Maceri, M.~Marino, and G.~Vairo.
\newblock A unified multiscale mechanical model for soft collagenous tissues
  with regular fiber arrangement.
\newblock \emph{Journal of biomechanics}, 43\penalty0 (2):\penalty0 355--363,
  2010.

\bibitem[Marino and Vairo(2013)]{Marino2013}
M.~Marino and G.~Vairo.
\newblock \emph{Multiscale Elastic Models of Collagen Bio-structures: From
  Cross-Linked Molecules to Soft Tissues}, pages 73--102.
\newblock Springer Berlin Heidelberg, Berlin, Heidelberg, 2013.

\bibitem[Marino et~al.(2015)Marino, Vairo, and Wriggers]{Marino2015}
M.~Marino, G.~Vairo, and P.~Wriggers.
\newblock Multiscale hierarchical mechanics in soft tissues.
\newblock \emph{Proceedings in Applied Mathematics and Mechanics}, 15\penalty0
  (1):\penalty0 35--38, 2015.

\bibitem[Garikipati et~al.(2008)Garikipati, G\"{o}ktepe, and
  Miehe]{Garikipati2008}
K.~Garikipati, S.~G\"{o}ktepe, and C.~Miehe.
\newblock Elastica-based strain energy functions for soft biological tissue.
\newblock \emph{Journal of the Mechanics and Physics of Solids}, 56\penalty0
  (4):\penalty0 1693--1713, 2008.

\bibitem[Holzapfel et~al.(2000)Holzapfel, Gasser, and Ogden]{Holzapfel2000}
G.~A. Holzapfel, T.~C. Gasser, and R.~W. Ogden.
\newblock A new constitutive framework for arterial wall mechanics and a
  comparative study of material models.
\newblock \emph{Journal of Elasticity}, 61:\penalty0 1--48, 2000.

\bibitem[Chan(1997)]{Chan1997}
L.~S. Chan.
\newblock Human skin basement membrane in health and autoimmune diseases.
\newblock \emph{Frontiers in Bioscience}, 2\penalty0 (July 15):\penalty0
  343--352, 1997.

\bibitem[Lavker(1979)]{Lavker1979}
R.~M. Lavker.
\newblock Structural alterations in exposed and unexposed aged skin.
\newblock \emph{Journal of Investigative Dermatology}, 73:\penalty0 59--66,
  1979.

\bibitem[Lagarrigue et~al.(2012)Lagarrigue, George, Questel, Lauze, Meyer,
  Lagarde, Simon, Schmitt, Serre, and Paul]{Lagarrigue2012}
S.~G. Lagarrigue, J.~George, E.~Questel, C.~Lauze, N.~Meyer, J.~M. Lagarde,
  M.~Simon, A.~M. Schmitt, G.~Serre, and C.~Paul.
\newblock In vivo quantification of epidermis pigmentation and dermis papilla
  density with reflectance confocal microscopy: variations with age and skin
  phototype.
\newblock \emph{Experimental dermatology}, 21\penalty0 (4):\penalty0 281--286,
  2012.

\bibitem[Sauermann et~al.(2002)Sauermann, Clemann, Jaspers, Gambichler,
  Altmeyer, Hoffmann, and Ennen]{Sauermann2002}
K.~Sauermann, S.~Clemann, S.~Jaspers, T.~Gambichler, P.~Altmeyer, K.~Hoffmann,
  and J.~Ennen.
\newblock Age related changes of human skin investigated with histometric
  measurements by confocal laser scanning microscopy in vivo.
\newblock \emph{Skin Research and Technology}, 8\penalty0 (1):\penalty0 52--56,
  2002.

\bibitem[Querleux et~al.(2009)Querleux, Baldeweck, Diridollou, de~Rigal,
  Huguet, Leroy, and Barbosa]{Querleux2009}
B.~Querleux, T.~Baldeweck, S.~Diridollou, J.~de~Rigal, E.~Huguet, F.~Leroy, and
  V.~H. Barbosa.
\newblock Skin from various ethnic origins and aging: an in vivo
  cross-sectional multimodality imaging study.
\newblock \emph{Skin Research and Technology}, 15\penalty0 (3):\penalty0
  306--313, 2009.

\bibitem[Kligman et~al.(1985)Kligman, Zheng, and Lavker]{Kligman1985}
A.~M. Kligman, P.~Zheng, and R.~M. Lavker.
\newblock The anatomy and pathogenesis of wrinkles.
\newblock \emph{British Journal of Dermatology}, 113:\penalty0 37--42, 1985.

\bibitem[Fenske and Lober(1986)]{Fenske1986}
N.~A. Fenske and C.~W. Lober.
\newblock Structural and functional changes of normal aging skin.
\newblock \emph{Journal of the American Academy of Dermatology}, 15\penalty0
  (4(1)):\penalty0 571--585, 1986.

\bibitem[Varani et~al.(2001)Varani, Spearman, Perone, Fligiel, Datta, Wang,
  Shao, Kang, Fisher, and Voorhees]{Varani2001}
J.~Varani, D.~Spearman, P.~Perone, S.~E. Fligiel, S.~C. Datta, Z.~Q. Wang,
  Y.~Shao, S.~Kang, G.~J. Fisher, and J.~J. Voorhees.
\newblock Inhibition of type i procollagen synthesis by damaged collagen in
  photoaged skin and by collagenase-degraded collagen in vitro.
\newblock \emph{Am J Pathol}, 158\penalty0 (3):\penalty0 931--42, 2001.

\bibitem[Ashworth et~al.(1999)Ashworth, Murphy, Rock, Sherratt, Shapiro,
  Shuttleworth, and Kielty]{Ashworth1999}
J.~L. Ashworth, G.~Murphy, M.~J. Rock, M.~J. Sherratt, S.~D. Shapiro, C.~A.
  Shuttleworth, and C.~M. Kielty.
\newblock Fibrillin degradation by matrix metalloproteinases: implications for
  connective tissue remodelling.
\newblock \emph{Biochemical Journal}, 340\penalty0 (Part 1):\penalty0 171--181,
  1999.

\bibitem[Humphrey and Rajagopal(2003)]{Humphrey2003}
J.~D. Humphrey and K.~R. Rajagopal.
\newblock A constrained mixture model for arterial adaptations to a sustained
  step change in blood flow.
\newblock \emph{Biomechanics and Modeling in Mechanobiology}, 2\penalty0
  (2):\penalty0 109--126, 2003.

\bibitem[Rouhi et~al.(2007)Rouhi, Epstein, Sudak, and Herzog]{Rouhi2007}
G.~Rouhi, M.~Epstein, L.~Sudak, and W.~Herzog.
\newblock Modeling bone resorbtion using mixture theory with chemical
  reactions.
\newblock \emph{Mechanics of Materials and Structures}, 2\penalty0
  (6):\penalty0 1141--1155, 2007.

\bibitem[Valent\'{i}n and Humphrey(2009)]{Valentin2009}
A.~Valent\'{i}n and J.~D. Humphrey.
\newblock Evaluation of fundamental hypotheses underlying constrained mixture
  models of arterial growth and remodelling.
\newblock \emph{Philosophical Transactions of the Royal Society A -
  Mathematical Physical and Engineering Sciences}, 367\penalty0
  (1902):\penalty0 3585--3606, 2009.

\bibitem[Valent\'{i}n and Holzapfel(2012)]{Valentin2012}
A.~Valent\'{i}n and G.~A. Holzapfel.
\newblock Constrained mixture models as tools for testing competing hypothesis
  in arterial biomechanics: Survey.
\newblock \emph{Mechanics Research Communications}, 29:\penalty0 126--133,
  2012.

\bibitem[Menzel and Kuhl(2012)]{Menzel2012}
A.~Menzel and E.~Kuhl.
\newblock Frontiers in growth and remodeling.
\newblock \emph{Mechanics Research Communications}, 42:\penalty0 1--14, 2012.

\end{thebibliography}
%
%
%
%
%
%

\end{document}